\documentclass[journal]{IEEEtran}

\usepackage{amsmath,amssymb,amsfonts}
\usepackage{algorithmic}
\usepackage{graphicx}
\usepackage{xcolor}
\usepackage{tikz}
\usepackage{algorithm}
\usepackage{ntheorem}
\usepackage{comment}
\usepackage{subfigure}
\usepackage{url}
\usepackage{caption}
\usepackage{comment}
\usepackage{booktabs}
\usepackage{multirow}
\newtheorem{theorem}{Theorem}
\newtheorem{lemma}[theorem]{Lemma}
\author{
    \IEEEauthorblockN{Abrar~Alali\IEEEauthorrefmark{1}\IEEEauthorrefmark{2},
    Stephan Olariu\IEEEauthorrefmark{1}, 
    Shubham Jain\IEEEauthorrefmark{3}}\\
    
    \IEEEauthorblockA{\IEEEauthorrefmark{1} Old Dominion University
    \\aalal003@odu.edu, olariu@cs.odu.edu}\\
    
    \IEEEauthorblockA{\IEEEauthorrefmark{2} Saudi Electronic University
    \\a.alali@seu.edu.sa}\\
    
    \IEEEauthorblockA{\IEEEauthorrefmark{3} Stony Brook University
    \\jain@cs.stonybrook.edu}
}

\title{\textit{ADOPT}: A system for Alerting Drivers to Occluded Pedestrian Traffic \thanks{Research funded by the NSF grant CNS-1951789 and CNS-2106594.}}
\begin{document}
\maketitle

\begin{abstract}
Recent statistics reveal an alarming increase in accidents involving pedestrians (especially children) crossing the street. A common philosophy of existing 
pedestrian detection approaches is that this task should be undertaken by the moving cars\footnote{In this work we refer to vehicles simply as cars.} 
themselves. In sharp departure from this philosophy, 
we propose to enlist the help of cars parked along the sidewalk to detect and protect crossing pedestrians. In support of this goal, we 
propose ADOPT: a system for Alerting Drivers to Occluded Pedestrian Traffic. 
ADOPT lays the theoretical foundations of a system that uses parked cars to: (1) detect the presence of a group of crossing pedestrians -- a crossing cohort;
(2) predict the time the last member of the cohort takes to clear the street; (3) send alert messages to those approaching cars that may 
reach the crossing area while pedestrians are still in the street; and, (4) show how approaching cars can adjust their speed, given several simultaneous crossing 
locations. 
Importantly, in ADOPT all communications occur over very short distances and at very low power. 
Our extensive simulations using SUMO-generated pedestrian and car traffic have shown the effectiveness of ADOPT in detecting and protecting crossing pedestrians.
\end{abstract}


\IEEEpeerreviewmaketitle

\section{Introduction and Motivation}

According to the National Highway Traffic Safety Administration, 6,301 pedestrians were hit and killed by drivers in the first half of 2021, a 17\% increase 
over 2020\footnote{The report accounted for the effects of COVID-19 pandemic.}~\cite{ghsa:2021a}.
A recent study~\cite{Yue2020-pk} concluded  that one of the main causes of crashes involving pedestrians is occlusion: the driver is 
unaware of the presence of pedestrians because some object %, typically a parked vehicle~\cite{nhtsa:2019b}, 
partially or fully occludes them. 
Therefore, detecting occluded pedestrians reliably and in a timely manner is key to promoting pedestrian safety. For comprehensive survey papers that
discuss the effectiveness of various approaches to improving pedestrian safety, we refer the reader to \cite{Mahdinia2022-kw} 
and \cite{El_Hamdani2020-vy}.

A glance at the recent literature reveals an increasing number of publications that leverage the on-board sensing and communication capabilities 
of present-day cars to heighten the drivers' awareness of surrounding pedestrians. 
On-board cameras, short- and long-range laser devices, and LiDAR are part of the arsenal of sensors employed to detect pedestrians~\cite{Gilroy2019-xl}. 
One common characteristic of these sensors is that they require Line-of-Sight (LoS) to enable 
detection~\cite{Ning2021-mp}. Because of this, partially or totally occluded pedestrians go undetected most of the time \cite{zhang2016far}.
In addition to the LoS challenge, weather and lighting conditions (e.g., glinting sun or other reduced visibility conditions)
are apt to thwart the ability of on-board sensors to detect pedestrians~\cite{Sun2020-ra, Combs2019-xy, combs2019automated}.

One approach for detecting pedestrians that does not require LoS involves leveraging wireless communication technologies such as
WiFi, DSRC, or short-range technologies such as Zigbee or Bluetooth. % to enable Vehicle-to-Pedestrian V2P communication~\cite{Sewalkar2019-dc}. 
While this approach mitigates the LoS challenge, it has serious scalability problems. Indeed, reporting pedestrians
within a large radio coverage area tends to be unreliable due to known impairments of radio transmission, various forms of interference,  
message propagation delays, and security concerns \cite{Rawat2014-ha}. 

Recently, researchers have suggested supplementing the
data collected by on-board sensors with information collected by pre-deployed roadside infrastructure~\cite{Sewalkar2019-dc}. 
Aligned with this idea, several projects have been implemented at signalized intersections to promote pedestrian safety~\cite{Grembek2019-te}. 
These projects involve installing 
sensors, cameras, and communication units in roadside infrastructure, such as  light poles, to detect pedestrians and to alert approaching 
cars \cite{Zhang2021-rr}. Although this approach enables the detection of occluded pedestrians, it focuses on signalized intersections, while 
the majority of pedestrian accidents tend to occur mid-block \cite{Tezcan2019-su,Li2021-ne,Zhang2017-xn}. Moreover, this approach is 
problematic since roadside infrastructure may not be available when needed~\cite{combs2019automated}. 

Recently, the potential of using parked vehicles as alternative road-side infrastructure in support of sensing and networking has been 
recognized by a number of researchers. For example,~\cite{Eckhoff2011-ln} and~\cite{Sommer2014-hp} 
suggested that by networking cars parked along the sidewalk, the task of relaying messages and other networking tasks can be significantly enhanced. 

More recently, it was suggested to aggregate information collected by several moving vehicles in an effort to enable a 
{\em collaborative perception} of surrounding pedestrians \cite{Sun2020-ra}. The idea is that when a moving vehicle has a LoS connection to a group of 
pedestrians, it alerts neighboring vehicles, making them aware of the presence of potentially occluded pedestrians. Alternatively, 
using Vehicle-to-Pedestrian (V2P) communications~\cite{Sewalkar2019-dc}, the pedestrians are 
alerted to the presence of approaching cars by pushing messages to their hand-held or wearable devices. Collaborative perception is a promising 
technology for enhancing 
pedestrian safety. However, in order to work effectively, the collaborative pedestrian detection methodology requires sophisticated sensing 
resources and communication protocols and may not scale well to a large number of pedestrians \cite{Loke2019-sn}. Furthermore, it is rather inaccurate 
since it relies on inaccurate GPS coordinates to estimate the distance between the pedestrian detected and the vehicle, not to mention the 
potentially high power consumption \cite{Yee2018-lj}. Most importantly, the alert messages sent via DSRC can easily overwhelm the 
network with unnecessary alerts \cite{Wu2013-wm}.

\subsection{Our contributions}

The underlying design philosophy of all the approaches mentioned above is that pedestrian detection should be undertaken by {\em moving cars} with all the 
complications that this entails. In sharp departure from this philosophy, we propose {\em ADOPT: a system for Alerting 
Drivers to Occluded Pedestrian Traffic} that runs on cars {\em parked} along the sidewalk. 

%as illustrated in Figure \ref{fig:sysconcept}.
%
%\begin{figure}[h!]
%    \centering
%    \includegraphics[width=0.48\textwidth]{Figures/SystemConcept4.png}
%    \caption{Illustrating ADOPT: Parked vehicles detect occluded crossing pedestrians and alert the approaching vehicles.}
%    \label{fig:sysconcept}
%\end{figure}

By design, ADOPT is a low-power system whose stated goal is to  alert approaching vehicles to the presence of 
possibly occluded pedestrians in the street. To implement this functionality, in ADOPT, all communications occur over very short distances 
and at very low power. 
Piezo-electric cells implanted in pedestrians' shoes power rudimentary radio transmitters that operate by closing a circuit as the pedestrian's shoe touches the
ground. The resulting signals (essentially, radio noise) allow low-power transceivers,
installed close to the four corners of parked cars, and integrated into the Controller Area Network (CAN) intra-vehicle network \cite{standard2003road}, 
to classify pedestrians as {\em on the sidewalk} or {\em in the street}. As long as the pedestrians stay on the sidewalk, ADOPT 
remains in a low-power vigilant state. When one or several pedestrians step into the street, ADOPT wakes up, locates the crossing pedestrians (a.k.a. crossing cohort), 
and estimates the time it takes the last member of the cohort to cross the street. By multi-hopping this information through a chain of parked cars along the sidewalk, 
ADOPT shares with the drivers of approaching 
cars (a) the location, on a digital map, of the crossing cohort ahead of them; and, (b) the estimated time it takes the cohort to cross the street.

To summarize, the main contributions of ADOPT are related to pedestrian safety and driver assistance. Specifically, this paper:
\begin{enumerate}
    \item provides the theoretical foundations of a low-power and infrastructure-free occluded pedestrian detection system;
    \item introduces a novel criterion for the classification of pedestrians as ``on the sidewalk'' or ``in the street'';
    \item offers a scheme for estimating the expected time it takes the crossing cohort to clear the street;
    \item provides an algorithm that allows approaching cars to adjust dynamically their speed, given several simultaneous crossing locations.
\end{enumerate}

The remainder of this work is structured as follows: Section~\ref{sec:related-work} offers a succinct survey of relevant recent work.
Section~\ref{sec:assumptions} establishes terminology and discusses system assumptions. A detailed discussion of how ADOPT works can be found in 
Sections \ref{sec:class+loc}--\ref{sec:approaching}. Specifically, Section \ref{sec:class+loc} offer the details of pedestrian classification and localization. 
Section \ref{sec:TTC} offers the details of the way ADOPT estimates the time it takes a cohort to cross the street. Next,
Section \ref{sec:propagation} discusses the details of propagating alert message to alert approaching cars. In Section~\ref{sec:approaching} we present an
algorithm that can be used by approaching cars to adjust their speed as a result of receiving information about one or several crossing cohorts.
The empirical evaluation of ADOPT spans Sections \ref{sec:sim-model} and \ref{sec:evaluation}. Specifically, Section \ref{sec:sim-model} introduces the
simulation model and the noise model assumed. Our comprehensive  simulation results are  presented and discussed in Section~\ref{sec:evaluation}. 
Finally, Section~\ref{sec:concl} offers concluding remarks and maps out directions for future work. 

\section{Relevant related work}\label{sec:related-work}

The main goal of this section is to offer a succinct overview of the latest  pedestrian detection systems and relevant emerging technologies.

\subsection{On-board LoS sensors}

The literature on pedestrian detection by vehicle's on-board sensors is quite vast 
\cite{Mahdinia2022-kw,Gilroy2019-xl,Ning2021-mp,zhang2016far,De_Nicolao2007-op,Kwon2016-xh,zhang2018occluded,Chen2019-pf,Palffy2019-dv,zhou2019discriminative,Brehar2021-fj,He2021-zs}. 
On-board cameras can detect pedestrians only if they fully appear in the view. However, 
if they are partially occluded, additional effort is needed to recognize them. 
In a series of papers, \cite{He2021-zs} have used machine learning to detect partially occluded pedestrians from an on-board camera. 
Convolutional neural networks were utilized in the detection of partially occluded pedestrians
~\cite{pang2019mask, zhang2018occluded, zhou2019discriminative}. \par
Sensor fusion approaches have been leveraged to detect occluded pedestrians. Recently, 
~\cite{Ding2022-ea}, \cite{Brehar2021-fj} uses on-board thermal infrared sensors to detect pedestrians in low lighting 
conditions.~\cite{Palffy2019-dv,Chen2019-pf} suggested fusing on-board camera input with LIDAR and 
thermal sensors to detect pedestrians in low-visibility conditions.\par
While on-board LoS sensors can detect partially occluded pedestrians, fully occluded pedestrians cannot be detected using the same sensors. Furthermore, relying on 
LoS sensors for occluded pedestrian detection must involve extensive vision algorithms that may not perform  the task in a timely manner. %Moreover, 
%even though the pedestrian can be detected when partially occluded, additional effort is required to predict the orientation and to identify whether the
%pedestrian has the intention to cross the street~\cite{Razali2021-ff}
\par

\subsection{V2P and P2V communications}
In order to mitigate the problem of occluded pedestrian detection with LoS sensors, Vehicle-to-Pedestrians (V2P) wireless communication was leveraged to detect 
the presence of pedestrians. WiFi, Zigbee, and Ultra-Wideband UWB were used in \cite{Ho2017-it, Dhondge2014-gq,Wang2021-mj,Hisaka2011-in,Zhang2019-iq} for 
V2P communication and alerting either drivers or pedestrians about anticipated collisions.\cite{Tahmasbi-Sarvestani2017-ym} have proposed 
a framework for V2P communications via DSRC units with the goal of alerting both the pedestrians and the vehicle to a possible collision. The fusion of the car's 
perception and V2P communications was leveraged by \cite{Merdrignac2017-lj}. Their solution relies on recognizing the occlusion by a 
moving vehicle and sending an alert to the pedestrian's device if detected via V2P communications.

Although V2P communications enhance the safety of occluded pedestrians by improving the detection rate, their main drawback is that they are usually power-hungry
and are apt to drain the battery of the pedestrian's device if used for extended periods of time. Yet another drawback is that V2P communications  rely on 
inaccurate GPS readings to determine the distance between the pedestrian and approaching vehicles. Additionally, V2P communications do not scale well when multiple
pedestrians are found in the street.

\subsection{Collaborative perception}
Collaborative perception in Vehicular Ad Hoc Networks (VANET) is used in pedestrian safety systems when a moving car fails to detect pedestrians using 
its on-board sensors and relies 
on remote sensors such as other cars' on-board cameras or street monitoring cameras to provide additional perception. \cite{Sun2020-dh} proposed a collaborative system that 
shares pedestrians' information when they are detected by another car's camera. 
The detecting car exchanges pedestrian location and speed with the blinded car(s). %which estimates the trajectory of the pedestrian. 
Their system has the advantage of enabling collaboration between cars to prevent pedestrian/car collisions. 
%However, the the detecting cars may also miss occluded pedestrians if parked cars obstruct the view. 

%Similarly, \cite{Ashraf2021-qy} utilized V2V communications to exchange warning messages about crossing pedestrians detected by on-board cameras. 
%Their system was able to identify occluded crossing pedestrians and warn neighboring cars using visible light communication. 
%The limitation of this work is that the occlusion scene needs to be detected and analyzed through image processing which requires extensive processing. 
%This limitation is challenging if implemented on parked cars. In addition, the detection car may also miss the pedestrians occluded by a parked car.

Several projects attempt to enhance pedestrian detection at intersections by installing cameras and ranging sensors on light poles or RSUs. Once 
the pedestrians are detected, approaching vehicles are alerted to their presence. For example, \cite{Islam2020-ol,Ben_Khalifa2020-ho,NOH2022103570} 
installed a camera in an RSU to detect pedestrians and to alert approaching vehicles.  \cite{Larson2020-qy} and \cite{Zhao2019-ei} suggested fusing 
input from installed cameras with thermal and LiDAR sensors to detect pedestrians at day and night. 
In lower cost mechanism, \cite{Pereira2020-zf} installed piezoelectric elements at the beginning and end of crossing lines to 
detect pedestrians and to alert approaching vehicles through RSUs. Similarly, \cite{Senart2008-dk} suggested placing wireless sensor units on the roadside to 
detect the presence of pedestrians and to send the detection information to a nearby RSU to share it with the incoming vehicles. The aforementioned approaches 
are expensive to deploy and, consequently, municipalities 
usually do not have the resources to install them at all locations. Besides, they only cover controlled intersections while many pedestrians are known 
to roads mid-block. Additionally, the alert messages are exchanged between RSU and approaching vehicles via DSRC which covers a large area causing 
network load and impairments of radio transmissions, various forms of interference, message propagation delays, and security concerns. \par

\subsection{Putting parked cars to work}
Parked cars have been used in VANET as relays for safety message dissemination in areas with low traffic density. 
As an example, \cite{Liu2011-nh} proposed the idea of using parked cars as RSUs to improve VANET connectivity. Similarly, \cite{Eckhoff2011-ln} showed that parked cars 
are useful to work as relay nodes in support of VANET communications. To increase road safety, \cite{Sommer2014-hp} proposed that parked cars communicate with their moving 
counterparts as relay nodes to increase safety in low-density areas by multi-hopping the Cooperative Awareness Messages. 
%We enabled parked cars in ADOPT to be cooperative in VANET and inform the approaching cars about detected pedestrians.

%\subsubsection {Parked cars as sensing resources}
Due to the vast panoply of their on-board sensors, parked cars can also be used as a sensing resource. 
\cite{Abdelhamid7018198} showed that parked cars  can be used as sensing resources and that the CAN bus can be used to provide power to the sensors. Parked cars have been 
also used for detecting  missing objects by \cite{Griggs2018-qt} by instrumenting cars with RFID readers to detect tags attached 
to missing objects, then share the detection information with administrative centers. One of the RFID detection approach's drawbacks is that it needs a secure 
communication medium to protect users' privacy. Undoubtedly, using parked cars as a sensing resource is beneficial in terms of providing accurate and timely 
information. 

\subsection{Energy harvesting wearables} 
Energy harvesters such as piezoelectric elements were used in shoes to activate sensors from mechanical energy generated during 
walking~\cite{Zhao2014-gi}. Moreover, \cite{Huang2018-qp}  developed a shoe module that can communicate with a smartphone held by the user. 
Their work showed that the power produced by a single shoe is sufficient to transmit data from a Bluetooth module to the smartphone. Such a module can be used to 
enable pedestrians to generate signals to be received by nearby receivers that operate on the same frequency as we proposed in this paper. Most 
importantly, they can transmit signals while consuming power from their motion rather than external batteries.

\subsection{Short-range RSS  localization} 
Received Signal Strength (RSS) ranging approaches are commonly used in wireless localization techniques for their simplicity and efficiency in terms of 
power consumption \cite{Zanella2016-im}. However, because of its limited accuracy, RSS-based ranging is used mostly as a coarse-grain indicator, 
especially in long-range localization efforts. As pointed out by \cite{Stoyanova2015-ro}, the accuracy of RSS localization is 
surprisingly good at short ranges (within 1-3 meters) which makes it a good candidate to be used in ADOPT.

\section{ADOPT: System Assumptions}\label{sec:assumptions}

The main goal of this section is to spell out the basic system assumptions that underlie ADOPT.

\begin{figure}[h!]
    \centering
    \includegraphics[width=0.5\textwidth]{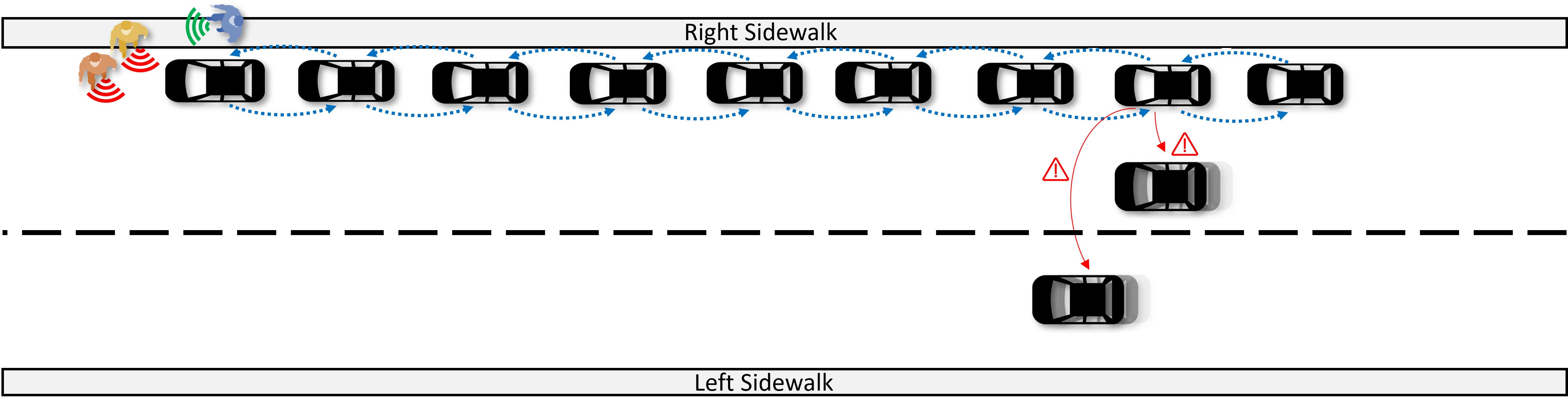}
    \caption{\em Illustrating ADOPT: Parked cars detect occluded crossing pedestrians and alert approaching cars.}
    \label{fig:sysconcept}
\end{figure}

ADOPT relies on detecting radio frequency (RF) signals transmitted at each step while a pedestrian is walking. Many {\em wearable devices}, including
smart-phones, wrist-bands, and shoes can detect human steps \cite{Seneviratne2017-kg}.. 
Since we aim to design a low-power and low-cost system, we focus on leveraging wearables that harvest energy from body motion. Specifically, we 
assume that the pedestrians (who might be children) are wearing shoes fitted with piezoelectric elements that provide a robust, lightweight, 
and inexpensive source of power for an in-shoe, battery-free, power generator \cite{Zhao2014-gi,Huang2018-qp}.
As already mentioned, when a pedestrian's shoe touches the ground, a circuit is closed and a rudimentary transmitter embedded in the pedestrian's show is 
activated for a fraction of a second.

We assume that the parked cars know their exact geographic location by interfacing with a digital map.
In support of detecting the presence of pedestrians, cars are fitted with four low-power radio transceivers placed on the front and rear 
axles on each side of the car. These transceivers detect and measure the strength of RF signals transmitted by pedestrians' shoes.
%Being close to the ground, we assume that there are no obstacles between the pedestrians' shoes and the transceivers.
The four transceivers mentioned above, are integrated into, powered by, and synchronized to the intra-vehicle CAN bus \cite{standard2003road}.

The cars parked along the sidewalk are assumed to be aware of the distance between their right transceivers and the sidewalk. This distance
can be estimated by using any on-board proximity sensor before stopping. 
%For later reference, we mention that empirical evidence seems to indicate that this distance is normally-distributed with a mean of about 40 cm.

Using well-known VANET protocols \cite{Balen2012-ux,vanet-book}, the cars parked along the sidewalk self-organize into a linear vehicular network. 
In this work, we refer 
to the resulting network as a {\em chain} of parked cars. Refer to Fig.~\ref{fig:sysconcept} for an illustration.
By consulting its on-board digital map, each car in the chain determines the width of the street and the speed limit.
It also identifies its position in the chain and the pseudonyms of its two adjacent neighbors. When a car is departing or joining the chain, a simple
maintenance operation is performed using well-known protocols discussed elsewhere \cite{Eckhoff2011-ln} or \cite{vanet-book}.

It is important to note that since in an on-street parking situation, adjacent parked cars are, typically, a short distance away from each other \cite{Aljohani2022-xj}, 
the tasks inherent to self-organization and maintenance 
of the chain of parked cars can be performed at low power using a suitable subset of transceivers.

We assume that when ADOPT is booted, there are no pedestrians in the street -- pedestrians, if any, are all on the sidewalk. This assumption is 
non-essential and is made for convenience only. Similarly, 
we assume that pedestrians step into the street from the sidewalk {\em in front} of a parked car. If the pedestrians step into the street
behind a given parked car, but in front of the next parked car in the chain, the latter will be responsible for undertaking the alerting actions. 
If there is no car parked behind, then the original parked car will undertake the alerting actions, as described in Section \ref{sec:propagation}, 
using its rear transceivers instead of the front ones.

Alert messages indicating the presence of a crossing cohort are propagated backward along the chain of parked cars, as illustrated in Fig.~\ref{fig:sysconcept}. 
As they pass cars in the chain of parked cars, approaching cars will be alerted, using low-power communications, to the reported location 
(or locations) of crossing pedestrians.

Although not specifically designed with self-driving cars in mind, ADOPT can be easily implemented 
to run on self-driving cars. When such a car receives an ADOPT ``Caution -- pedestrian in the street'' message, it complies 
by reducing its speed without delay using a control system such as the system proposed by \cite{SAM201617}. However, at the moment, self-driving cars 
are not very common, and so ADOPT alerts human drivers by displaying appropriate messages on a dashboard digital map, attempting to 
minimize distraction. For this purpose, we assume an ADOPT app that is running in approaching cars. The app determines the speed reduction necessary to avoid 
collision with the crossing cohorts and displays the suggested speed on the dashboard. 

\section{Pedestrian Classification and Localization}\label{sec:class+loc}

The first task that parked cars need to undertake is to determine if there are pedestrians in the street. If no pedestrians are in the street all is well. 
Otherwise, the pedestrians have to be accurately localized and their crossing time estimated. Once this information is in hand, approaching cars are 
alerted to the presence of the crossing cohort.
The main goal of this section is to provide the details of pedestrian classification and localization. 
%The task of alerting approaching cars to the presence of a crossing cohort will be detailed in Section \ref{sec:propagation}.

%\begin{figure}[ht]
%    \centering
    %\includegraphics[width=0.3\textwidth]{Figures/SystemOverViewSECON.pdf}
 %   \caption{\em A high-level view of ADOPT.}
 %   \label{fig:overviewBlock}
%\end{figure}

\subsection{Pedestrian Detection} \label{sec:detection}

Recall that, as mentioned in Section \ref{sec:assumptions}, with each step the pedestrians' shoes generate, for a fraction of a second,  an RF signal 
with a known power $T$ and a known frequency $f$ \cite{Huang2018-qp}. 
When a pedestrian walks near a parked car, the transceivers in the car receive the signals generated by the pedestrian's shoes. We calculate 
the Received Signal Strength $RSS(Rx)$ at a generic transceiver $Rx$ using the Free Space propagation model \cite{lo2013antenna}:
\begin{equation}\label{Eq:freespace}
      RSS(Rx) = \frac{T\gamma}{\delta_{Rx}^2},
\end{equation}
where $T$ is the power of the transmitted signal, $\gamma$ is an environmental constant, and $\delta_{Rx}$ is the distance between the pedestrian 
who transmits the signal and the transceiver $Rx$.

\subsection{Pedestrian Classification}\label{subsec:classification}

An important task that ADOPT undertakes is the {\em binary} classification of pedestrians: {\em on the sidewalk} or {\em in the street}. 
We begin by stating and proving a technical result, of an independent interest, that provides a simple criterion for classifying pedestrians.
\begin{theorem}\label{thm:fundamental}
Consider a pedestrian carrying a transmitter and consider a car parked parallel to the edge of the sidewalk. Assume the car is provided with transceivers, L and R, placed at the 
front corners of the car. Then, the pedestrian is walking parallel to the edge of the sidewalk if, and only if, the difference between the reciprocals of the 
received signal strengths $RSS(R)$ and $RSS(L)$ at R and L, respectively, is a constant:
\begin{equation}\label{RSS}
\frac{1}{RSS(L)} - \frac{1}{RSS(R)} = const.
\end{equation}
\end{theorem}

\begin{IEEEproof}
Referring to Fig.~\ref{fig:RSS}, consider the horizontal line through the points L and R and let P be the position of the pedestrian. 
Assume, without loss of generality, that the projection, Q, of P onto this line lies to the right of R. 
Denote by $d$ the length of the segment PQ, by $w$ the length of the segment LR\footnote{It is useful to think of $w$ as the width of the car.}, and by $x$ 
the length of the segment RQ.

\begin{figure}[t]
    \centering
    \includegraphics[width=0.35\textwidth]{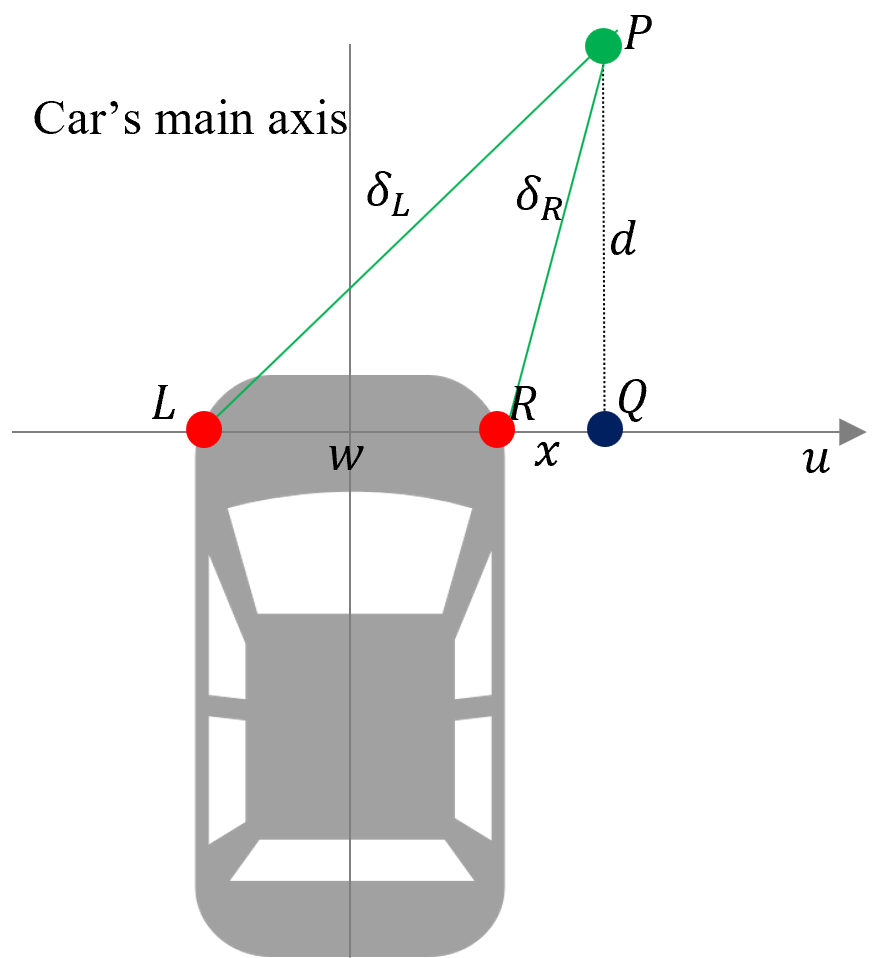}
    \caption{\em Illustrating the proof of Theorem \ref{thm:fundamental}}
    \label{fig:RSS}
\end{figure}

Further, let $\delta_R$ and $\delta_L$ be the Euclidean distance between the points P and R, and P and L, respectively.
By applying the Pythagorean theorem to triangles LPQ and RPQ we write
    \begin{equation*}
        \begin{cases}
          \delta_L^2 = (w+x)^2+d^2 & \text{and}\\
          \delta_R^2 = x^2+d^2 & \\
        \end{cases}
    \end{equation*}
    and consequently,
    \begin{equation}\label{eq:diff}
        \delta_L^2 - \delta_R^2 = (w+x)^2+d^2-x^2-d^2 = w^2+2wx.
    \end{equation}
    
On the other hand,
    \begin{equation*}
        \begin{cases}
          RSS(R) = \frac{T ~ \gamma}{\delta_R^2} & \text{and}\\
          RSS(L) = \frac{T ~ \gamma}{\delta_L^2} & \\
        \end{cases}
    \end{equation*} 
implying that
\begin{equation}\label{RSS-1}
\frac{1}{RSS(L)} - \frac{1}{RSS(R)} = \frac{ w^2+2wx}{T ~ \gamma}.
\end{equation}

Now, notice that in the right-hand-side of (\ref{RSS-1}), $w,\ T$ and $\gamma$ are constants. With this in mind:
\begin{itemize}
\item If the pedestrian walks parallel to the edge of the sidewalk) then $x$ remains constant and
so the left-hand-side is a constant;
\item Conversely, if $\frac{1}{RSS(L)} - \frac{1}{RSS(R)}= const$ then, since $T,\ \gamma$ and $w$ are constants,  $x$ itself must be a constant 
and so the pedestrian must walk parallel to the edge of the sidewalk.
\end{itemize}
This completes the proof of the theorem.
\end{IEEEproof}
\vspace*{2mm}

\noindent
{\bf Discussion}~ Equation (\ref{RSS}) establishes a one-to-one correspondence between a subset of real numbers and the set of lines 
in the plane parallel to the edge of the sidewalk. This is to say, to each line in the plane parallel to the edge of the sidewalk, 
there corresponds a unique real number and, 
conversely, to each real number there corresponds exactly one line in the plane parallel to the edge of the sidewalk.

By Theorem \ref{thm:fundamental}, all the points on the edge of the sidewalk feature the same constant $c_0$. Referring, again, to Fig.~\ref{fig:RSS}, imagine a 
transmitter placed at  Q. Here, $\delta_L=w+z,\ \delta_R=z$ and so, by (\ref{RSS-1}), 
\begin{equation}\label{c0}
c_0 = \frac{ w^2+2wz}{T ~ \gamma},
\end{equation}
where, recall,  $z$ is the  distance between the car and the sidewalk. 
%Empirical evidence seems to indicate that $z$ is normally-distributed with a mean of about 40cm.

The one-to-one correspondence discussed above is a very useful property because in order to  determine if a pedestrian is on the sidewalk, 
all we have to do is to evaluate the left-hand side of (\ref{RSS}) and to compare the result to the value of $c_0$ from  (\ref{c0}). 
This, in fact, is tantamount to a coarse-grain binary localization of pedestrians: on the sidewalk, or else in the street.

\subsection{Pedestrian Localization}\label{subsec:localization}

\begin{figure}[!h]
    \centering
    \includegraphics[width=0.3\textwidth]{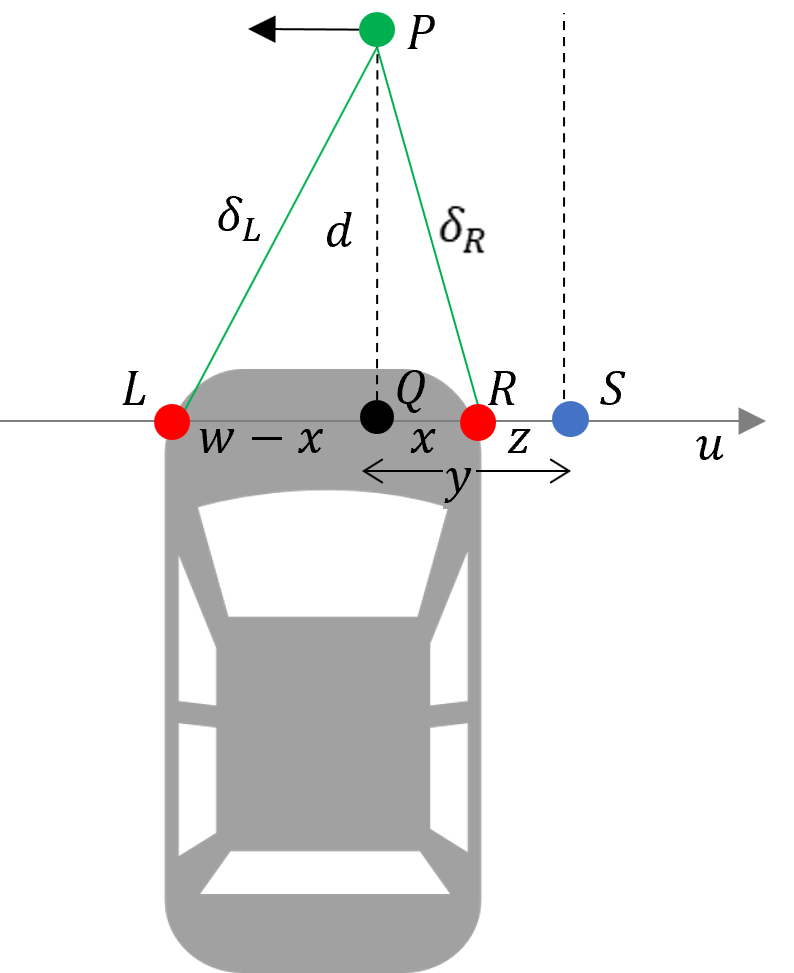}
    \caption{Illustrating the notation for the proof of Lemma \ref{lem:y}.}
    \label{fig:y}
\end{figure}

Pedestrians that are classified as {\em in the street} have to be localized more accurately, as we are about to explain.
Referring to Fig.~\ref{fig:y}, consider a pedestrian P and let point Q be the projection of P onto the line
determined by L and R.\footnote{It is useful to think of L and R as the left and transceivers at the front of the car.} We assume, without loss of generality,
that Q lies between L and R. We assume that the pedestrian crosses
the street by walking in a straight line parallel to the front of the parked car (i.e., perpendicular to the sidewalk). Let $d$ be the length of the line segment 
PQ (that is, the vertical distance between the location of the pedestrian and the line LR. 
Further, let $w$ be the width of the parked car, let $x$ be the length of the line segment QR, let $z$ be the distance from R to the sidewalk, and write
$y = x+z$.

Now, elementary geometry confirms that $y$ has the following expression
\begin{lemma}\label{lem:y}
\begin{equation}\label{eq:y}
y = \frac{w}{2} +z - \frac{T \gamma}{2w} \left [ \frac{1}{RSS(L)} - \frac{1}{RSS(R)} \right ].
\end{equation}
\end{lemma}

\begin{IEEEproof}

By using the Pythagorean theorem in the triangles PQL and PRQ we can write:
    $$
    \begin{cases}
          \delta_L^2 = (w-x)^2+d^2 & \text{and}\\
          \delta_R^2 = x^2+d^2 & \\
    \end{cases}
    $$
    and so:
    \begin{equation}\label{eq:diff2}
        \delta_L^2 - \delta_R^2 = (w-x)^2-x^2 = w^2-2wx.
    \end{equation}
    From~\eqref{Eq:freespace},
   $$
   \begin{cases}
   \delta_L^2 = \frac{T\gamma}{RSS(L)} &\text{and} \\
   \delta_R^2 = \frac{T\gamma}{RSS(R)}&\\
   \end{cases}
   $$
   by plugging these values into \eqref{eq:diff2}, we obtain
   $$
    \frac{T\gamma}{RSS(L)}- \frac{T\gamma}{RSS(R)} = w^2-2wx.
   $$
   Solving for $x$ yields
   \begin{equation}
       \begin{split}
       x &= \frac{w^2}{2w}-\frac{T\gamma}{2w} 
       \left[ \frac{1}{RSS(L)}-\frac{1}{RSS(R)}  \right]\\
       &=
       \frac{w}{2}-\frac{T\gamma}{2w} 
       \left[ \frac{1}{RSS(L)}-\frac{1}{RSS(R)}  \right]
       \end{split}
   \end{equation}
   Finally,
    \begin{equation}
       \begin{split}
       y & = x+z\\
       &= \frac{w}{2} +z - \frac{T \gamma}{2w} \left [ \frac{1}{RSS(L)} - \frac{1}{RSS(R)} \right ].
       \end{split}
   \end{equation}
   as claimed.
\end{IEEEproof}

Now, referring again to Fig.~\ref{fig:y}, it is clear that since the pedestrians are crossing the street in a direction perpendicular to the sidewalk, in order 
to specify the location of a crossing pedestrian, it suffices to specify the $y$
value, as above, along with the value of $d$, the vertical distance to the front of the parked car. 

In order to determine $d$, we compute the area of the triangle PLR
in two different ways:
\begin{itemize}
\item First, evidently, $Area(PLR) = \frac{w \cdot d}{2}$;
\item Second, writing  $p = \frac{\delta_L+\delta_R+w}{2}$, the same area can be expressed as 
$$
Area (PLR) = \sqrt{p(p-\delta_R)(p-\delta_L)(p-w)}
$$
\end{itemize}

Consequently, 
$$
\frac{w \cdot d}{2} = \sqrt{p(p-\delta_R)(p-\delta_L)(p-w)},
$$
which, upon solving for $d$, yields:
\begin{equation}\label{eq:d}
d = \frac{2}{w}\sqrt{p(p-\delta_R)(p-\delta_L)(p-w)}.
\end{equation}

\section{Estimating the time to cross}\label{sec:TTC}

Recall that a crossing cohort is a group of pedestrians crossing together at the same location. Instead of dealing with each member of the cohort
individually, we only concern ourselves with the {\em tail} of the cohort, defined as the last pedestrian in the cohort. It is clear that if the tail of the cohort
has crossed safely, then all pedestrians in the cohort have crossed safely, too.
It is important to note that ADOPT is {\em privacy-aware} as we are only interested in the location of the tail of the cohort, and not in the actual person that
happens to be the last in the cohort.

We now  define the tail of a crossing cohort more formally. 
Recall that the parameter $y$, defined in Lemma \ref{lem:y} keeps track of the current distance of a crossing pedestrian to the sidewalk she just departed. 
Formally, then, at each moment in time, the tail of the crossing cohort is the pedestrian with 
\begin{equation}\label{eq:tail}
\min \{y|\ {\rm pedestrian\ classified\ as\ in\ the\ street}\}.
\end{equation}
Notice that the tail may change dynamically either because new pedestrians joined the cohort or, simply, because some folks in the cohort 
walk faster than others. ADOPT updates the tail of the crossing cohort every second.

To manage the tail of a cohort, ADOPT needs to process signals from several pedestrians simultaneously. To accomplish this, the transceivers in the parked car
sample frequency $f$ on which the pedestrians' shoes are transmitting signals, $m$ times per second. Conceptually, this means that each second 
is partitioned into $m$ slots.  Assume that each pedestrian takes one step per second, and that each step
generates one transmission. For all practical purposes, this transmission occurs, randomly,
in one of the $m$ slots discussed above. Now, suppose that there are $k,\ (k \geq 1)$, pedestrians in the crossing cohort. We have just set up a ``balls-into-bins''
model involving $k$ balls and $m$ bins.  If two or more pedestrians are transmitting in the same time slot, a collision occurs and the outcome cannot be disambiguated.

We are interested to assess the expected number of ``clear'' transmissions, where one single transmission occurs in a given time slot.
For arbitrary $i,\ (1 \leq i \leq k)$, let $X_i$ be the indicator random variable that takes on the value 1 if, in a given second, slot $i$ sees a clear
transmission and $0$ otherwise. It is easy to see that
\begin{equation*}
\Pr[X_i =1] =\left ( 1-\frac{1}{m} \right )^{k-1}
\end{equation*}
and that the expected number, $E[M]$, of clear transmissions is
\begin{eqnarray*}
	E[M] &=& E[X_1 + X_2 + \cdots + X_k] \nonumber \\
	     &=& E[X_1] + E[X_2] + \cdots + E[X_k] \\
             &=& \Pr[X_1] + \Pr[X_2] + \cdots + \Pr[X_k] \\
             &=& k \left ( 1-\frac{1}{m} \right )^{k-1}.
\end{eqnarray*}

As an illustration, if a given cohort were to contain $k=8$ pedestrians, and assuming that frequency $f$ is sampled 50 times per second, we would expect to see
$E[M]= 8 \times \left ( 1-\frac{1}{50} \right )^{7} = 8 \times (\frac{49}{50})^7 \approx 6.9$ clear transmissions each second.

With this in mind, consider the time ruled into seconds and assume that in second $t$, the tail of the current crossing cohort was located 
at $y(t)$ and its current speed, $v(t)$, has been estimated. The {\em remaining time to cross}  at time $t$, $\Delta(t)$, can be estimated as
\begin{equation}\label{delta-t}
\Delta(t) = \frac {W-y(t)} {v(t)},
\end{equation}
where $W$ is the width of the street that the pedestrians are crossing.

We need to show how these parameters are updated in the next second, $t+1$. We begin by identifying all clear transmissions in second $t+1$ and, using
equation (\ref{eq:tail}), we obtain the location, $y(t+1)$, of the current tail. We distinguish between the following cases:  

\noindent
{\bf Case 1: $y(t) < y(t+1).$}

Evidently, in this case, the new tail is closer to the opposite sidewalk. It follows that no new pedestrian has joined the cohort in this
second. In this case, it is natural to update the cohort parameters as follows:
    \begin{itemize}
       \item $v(t+1) = y(t+1) - y(t)$ m/s;
       \item $\Delta(t+1) = \frac {W-y(t+1)} {v(t+1)}$.
    \end{itemize}
\vspace*{1.5mm}

\noindent
{\bf Case 2: $y(t) > y(t+1).$}

In this case, it is clear that one or more  pedestrians have joined the cohort and, consequently, the new tail
must be selected from among the pedestrians who have just joined the cohort.  There is a complication: we cannot update the speed of the tail, 
because the tail is new. Instead, we assign to the new tail, tentatively, the average crossing speed. The cohort parameters are updated as follows:
    \begin{itemize}
       \item $v(t+1) = v_0$,  where $v_0$ is an estimate of the average pedestrian speed;
       \item $\Delta(t+1) = \frac{W-y(t+1)}{v(t+1)}$.
    \end{itemize}

\section{Safety Zone and Message Propagation}\label{sec:propagation}

ADOPT involves two types of messages, each with its own semantics:
\begin{itemize}
\item {\em Alert messages:} are sent by the parked car that detects a crossing cohort with the intention of establishing a {\em Safety Zone}, as we are about to describe;
\item {\em Caution messages:} are messages sent by the parked cars in the Safety Zone to alert approaching (i.e. moving) cars to the presence of crossing pedestrians.
\end{itemize}

These to types of messages will be discussed in Subsections \ref{subsec:alert} and \ref{subsec:caution}, respectively.

\subsection{Alert messages}\label{subsec:alert}
\begin{figure}[ht]
    \centering
    \includegraphics[width=0.5\textwidth]{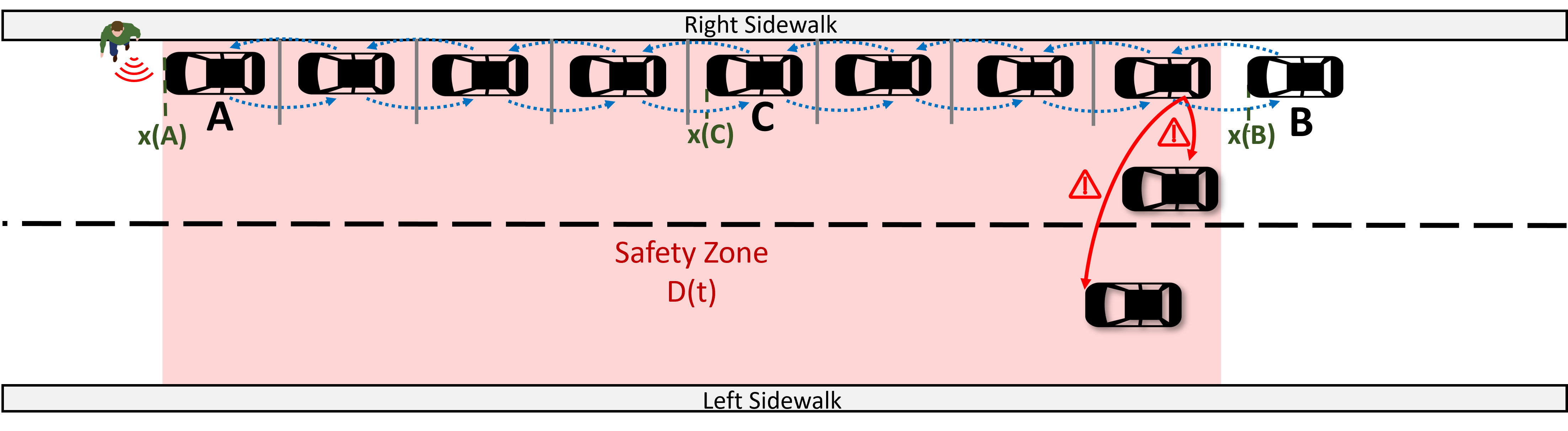}
    \caption{\em Parked cars in the Safety Zone propagate alert messages within the estimated propagation distance to alert approaching cars.}
    \label{fig:parkedcarchain2}
\end{figure}

Referring to Figure~\ref{fig:parkedcarchain2}, assume, without loss of generality, that car A, the first car 
in the chain of parked cars, detects a crossing cohort at time $t$. Proceeding as discussed in Section \ref{sec:TTC}, car A estimates the remaining
time, $\Delta(t)$ (see (\ref{delta-t})), it takes the tail of the cohort to cross the street. Using this information, car A determines the distance, $D(t)$, 
the alert messages will have to be propagated along the chain as follows:

\begin{equation}\label{eq:sz}
    D(t) = \left [\Delta(t)+r \right ] \cdot v_{max},
\end{equation}
where $r$ is the {\em driver reaction} time, estimated to be between 1.24 seconds and 2 seconds \cite{Koppa}. If self-driving cars are considered, then $r$ is set 
to zero, since the driver reaction time is a human factor and does not affect self-driving cars. Finally, car A propagates the alert message containing: 
its own location, $x(A)$, on the digital map, the current time, $t$, and the Distance-to-Live $D(t)$ of the alert message in meters. 

Referring to Fig.~\ref{fig:parkedcarchain2} again, the area of length $D(t)$, defined in~\eqref{eq:sz}, starting at car A and running down the chain of parked cars is called 
the {\em Safety Zone} associated with the crossing cohort detected by car A. 
Each parked car in the chain of parked cars, upon  receiving the alert message originated by car A, compares its own location with that of A and determines 
if the distance from A is smaller than or equal to $D(t)$. If so, it marks itself as being in the Safety Zone and propagates the alert message further along the chain.

As an illustration, in  Fig.~\ref{fig:parkedcarchain2}, assume that car C has just received the alert message from the previous car in the chain.
Let $x(A)$ and $x(C)$ be the x-coordinates of A and C, respectively. If $|x(A) - x(C)| \leq D(t)$ then C remembers that it belongs to the Safety Zone and 
propagates the message further down the chain. Continuing in this way,
the alert message will reach, eventually, car B. When car B receives the alert message, it finds that $|x(A) - x(B)| > D(t)$, and so it discards the message. 
As illustrated in Fig.~\ref{fig:parkedcarchain2}, car B is not in the Safety Zone.

\subsection{Caution messages}\label{subsec:caution}

Each parked car inside the Safety Zone is tasked with broadcasting a ``Caution -- pedestrians in the street'' (Caution, for short) message to approaching cars.
This broadcast must be done at low power, as passing cars are a short distance away from parked cars. The Caution message contains:
\begin{itemize}
\item the location $x(A) +d$ of the crossing cohort, where $d$ was computed in (\ref{eq:d});
\item the time, $t+\Delta(t)$, at which the cohort is expected to have crossed the street.
\end{itemize}

\begin{figure}[ht]
  \centering
  \includegraphics[width=0.5\textwidth]{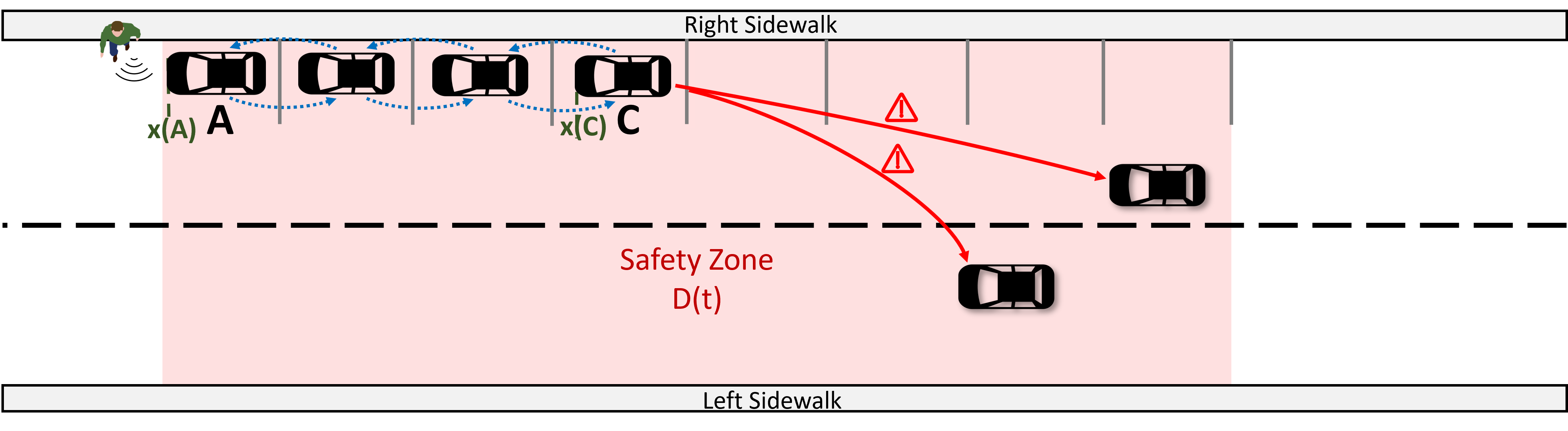}
  \caption{\em The Last car in the chain delivers the Caution  message to the approaching car via DSRC if the chain length is less than $D(t)$.}
  \label{fig:parkedcarchain3}
\end{figure}

In the unlikely event where many cars in the chain of parked cars depart, leaving big  gaps in the chain, as shown in Fig.~\ref{fig:parkedcarchain3}, 
the last car in the chain uses its DSRC transmitter to broadcast the Caution  message to inform  approaching cars of the location of the crossing cohort. 
As an illustrated in Fig.~\ref{fig:parkedcarchain3}, car C with $|x(A) - x(C)| < D(t)$ does not detect any car behind it. Hence, car C will use 
its DSRC transmitter to broadcast the "Caution"  message at a distance of $D(t)-x(C)$, where $x(C)$ is the location of C.

%The ``Caution -- pedestrians in the street'' message contains:
%\begin{itemize}
%\item the location $x(A) +d$ of the crossing cohort, where $d$ was computed in (\ref{eq:d});
%\item the time, $t+\Delta(t)$, at which the cohort is expected to have crossed the street.
%\end{itemize}

In the next second, $t+1$, a new estimate $D(t+1)$ is made as discussed in Section \ref{sec:TTC}. If $D(t+1) > D(t)$ then an updated alert message is sent with 
Distance-to-Live $D(t+1)$. Otherwise, no action is needed.

\begin{comment}
The ``Caution'' message must be broadcast as soon as the parked car receives the alert message originated with car A.
It is important to determine the time interval during which a parked car that has received an alert message should broadcast 
the ``Caution'' message to possible passing cars. Referring to Figure ~\ref{fig:parkedcarchain}, assume that car B has received the alert message at time
$\delta_B$. Car B will transmit the message in a very short time, i.e. in the interval 
$[\delta_B, \delta_B + \frac{|x(B) - x(B)|}{v_{limit}]} = [\delta_B, \delta_B]$. This is because cars that reach B beyond Consider an arbitrary car C with 
coordinates $x(C)$ that received the alert message originating at car A at time $\delta_C$ will transmit the ``Caution''  message in the 
interval $[\delta_C, \delta_C + \frac{|x(C)-x(B)|}{v_{limit}}]$, where $\v_{limit}$ is the speed limit. This is because  
In particular, car B will transmit the message in a very short time, i.e. in the interval $[\delta_B, \delta_B]$. Ideally, the ``Caution''' message 
should be broadcast in a staggered fashion to avoid collisions.
\end{comment}

\section{How Approaching Cars Determine a Safe Speed}\label{sec:approaching}

The main goal of this section is to show how approaching cars, alerted to the presence of crossing cohorts, adjust their speed in such a way that they avoid
colliding with the crossing pedestrians.

Our approach is novel and is based of a new way of looking at the time-space diagram (see the Appendix for a refresher).

\begin{figure}[h!]
    \centering
    \includegraphics[width=0.3\textwidth]{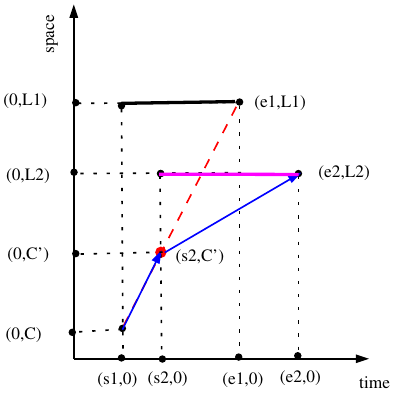}
    \caption{\em Illustrating the computation of the safe average speed in the case of two crossing cohorts.}
    \label{fig:2-cohorts}
\end{figure}

Referring to Fig.~\ref{fig:2-cohorts}, consider a car moving North-bound along a street. At time $s1$, the car receives a ``Caution -- pedestrians in the street'' 
message alerting it to a crossing cohort at location $(0,L1)$. Assume that the location of the car at time $s1$ is $(s1,C)$ and that the cohort 
will finish crossing the street at time $e1$. 

Proceeding as indicated in Subsection \ref{subsec:cohort} of the Appendix, the approaching car computes the maximum safe speed:
\begin{equation}\label{eq:vsafe}
	v_{safe} = \min \left \{ v_{max} , \frac{L1-C}{e1-s1} \right \},
\end{equation}
where $v_{max}$ is the speed limit along the street.

Traveling at this safe speed North-bound, the approaching car receives, at time $s2$, a second ``Caution -- pedestrians in the street'' message alerting it to the
presence of a second crossing cohort. This cohort crosses the street at location $(0,L2)$ and will finish crossing at time $e2$. How should the car 
change its speed to avoid an accident?

In order to answer this question, the first task is to determine the location, $(s2,C')$ of the car at time $s2$. This can be done by noting 
that $v_{safe}$ in equation (\ref{eq:vsafe}) can we written as
$$
\frac{L1-C}{e1-s1} = \frac{C'-C}{e1-s1}.
$$

Solving for $C'$ yields:
$$ %\begin{equation}\label{C'}
C' = C+ \frac{(L1-C)(s2-s1)}{e1-s1}.
$$ %\end{equation}

With $C'$ firmly in hand, the car updates the current safe speed $v_{safe}$ in (\ref{eq:vsafe}) as follows:
\begin{eqnarray}\label{safe2}
v_{safe} &=& \min \left \{ v_{safe}, \frac{L2-C'}{e2-s2} \right \} \nonumber \\
         &=& \min \left \{ v_{max}, \frac{L1-C}{e1-s1}, \frac{L2-C'}{e2-s2} \right \}.
\end{eqnarray}

The justification of (\ref{safe2}) is simple. The car needs to select the largest average safe speed and this is the smallest of the slopes of
the lines segments determined by the points $(s2,C')$ and $(e1,L1)$ on the one hand, and the points $(s2,C')$ and $(e2,L2)$ on the other.

At time $e1$, the car realizes that the first cohort has finished crossing the street and will adjust its speed again:
\begin{eqnarray}\label{safe3}
v_{safe} &=& \min \left \{ v_{max}, \frac{L2-C'}{e2-s2} \right \}.
\end{eqnarray}

In Fig. \ref{fig:2-cohorts}, 
$$\min \left \{ v_{max}, \frac{L2-C'}{e2-s2} \right \} = \min \left \{ v_{max}, \frac{L1-C}{e1-s1}, \frac{L2-C'}{e2-s2} \right \},$$ 
and, consequently, the car continues driving at the same speed.

Next, at time $e2$, the car realizes that the second cohort has finished crossing and so it will adjust its speed again:
\begin{eqnarray}\label{safe4}
v_{safe} &=& \min \left \{ v_{max} \right \} = v_{max},
\end{eqnarray}
essentially, reverting to the maximum allowed speed.

The same procedure is then continued, exactly as described, should other ``Caution -- pedestrians in the street'' be received by the car in the future.

\section{Simulation Model}\label{sec:sim-model}
\begin{figure}[ht]
    \centering
    \includegraphics[width=0.5\textwidth]{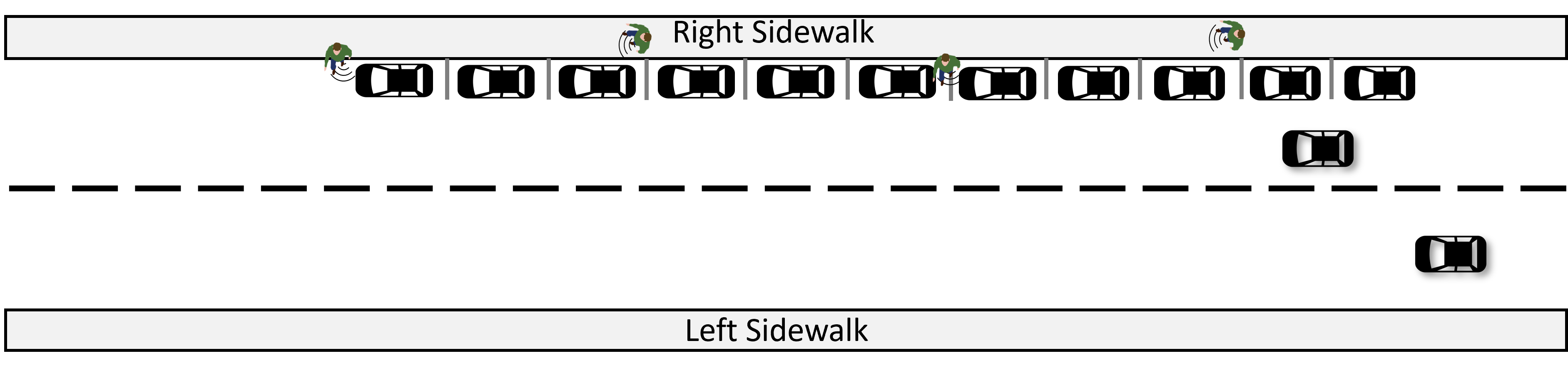}
    \caption{\em An instance of the ADOPT simulation model.}
    \label{fig:sumo}
\end{figure}

We validated the theoretical findings of ADOPT by testing them on simulated traffic data. For this purpose, we used SUMO (Simulation of Urban MObility) ~
\cite{SUMO2018} to generate pedestrian and car traffic data for the ADOPT simulation model illustrated in Fig.~\ref{fig:sumo}. 

Our simulation model consists of a {\em one-way} street with a road-side parking lane on the right side of the street and sidewalks on  both sides. 
Approaching cars enter at the end of the street and exit from the opposite end. 
In our simulation model, the pedestrians are not restricted to crossing at intersections but, indeed, they may cross {\em midblock} a prevalent 
behavior \cite{Tezcan2019-su}. 
%We dedicated the first car before each crossing area to detect crossing pedestrians and disseminate the alert message.
\vspace*{2mm}

\noindent
For the reader's convenience, we summarize the simulation parameters in Table \ref{tab:simulationSettings}.

\begin{table}[ht]
\caption{\em A summary of simulation parameters}
\begin{center}
\begin{tabular}{ll}
\hline
\textbf{Parameter}                         & \textbf{Value} \\
\hline
RF signal frequency $f$               & 2.4 GHz      \\
Transmission Power $T$                    & 2 mW      \\
Street width $W$                          & 12.8 m        \\
Distance from R transceiver to the sidewalk $z$ & 0.4 m \\
Number of detected pedestrians/sec $[\mu, \sigma]$ &     [1.8,1.4]     \\ 
Pedestrians speeds $[\mu, \sigma]$  & [1.15, 0.13]  m/sec    \\
Average speed $v_0$ & 1.2  m/sec    \\
Number of crossing pedestrians/sec $[\mu, \sigma]$  & [0.11, 0.32]         \\
Street speed limit  $v_{max}$ & 15 m/sec         \\
Simulation step length                 & 1 sec         \\
Simulation steps                  & 3600 sec (1 hour)       \\
\hline
\end{tabular}
\label{tab:simulationSettings}
\end{center}
\end{table}
\vspace*{0.5mm}
The data we collected from SUMO are as follows: street and sidewalk dimensions, pedestrians' locations, pedestrians' speeds, parked cars' locations and dimensions, 
and approaching cars' speed and location at each time step. Data collected from SUMO is our ground truth, i.e., ``actual'', information about pedestrians and cars 
traffic. 

We modeled pedestrians' radio signals by setting a constant transmission power equal to 2 mW transmitted from their actual locations. We then obtained the actual 
Euclidean distances $\delta_{Rx}$ between the pedestrian locations and the four transceivers located at the four corners of cars.
%We removed the samples where the distance from the nearest corners is greater than the assumed transmission range (3 meters). 
%\textcolor{red}{Samples that have distances less than 3 meters were used in \eqref{Eq:freespace} to obtain $RSS(Rx)$ at the two nearest transceivers because the transmission 
%range of signals produced by pedestrians' shoes was assumed to be 3 meters}. 

\subsection{Modeling RSS Noise}\label{subsec:noise-model}

In real-life situations, the Received Signal Strength ($RSS$) measurements experience random variations due to hardware-based parameters such as thermal noise, 
or due to the natural behavior of the signal as it is reflected by the ground~\cite{Stoyanova2010-hp}. Being close to the ground, we assume that there are no 
obstacles between the pedestrians' shoes and the transceivers.

\begin{figure}[ht]
\centering
    \subfigure[]{\includegraphics[width=0.3\textwidth]{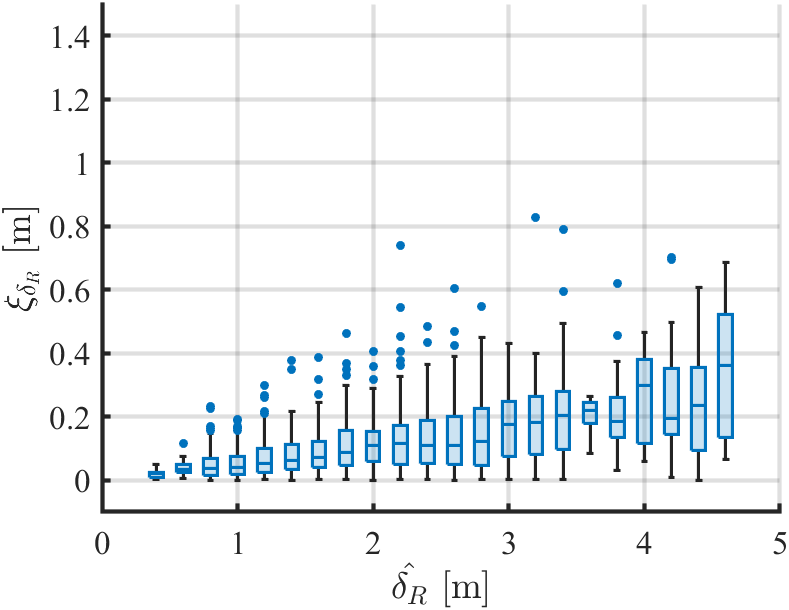}\label{fig:dr}}
    \hspace{3em}
    \subfigure[]{\includegraphics[width=0.3\textwidth]{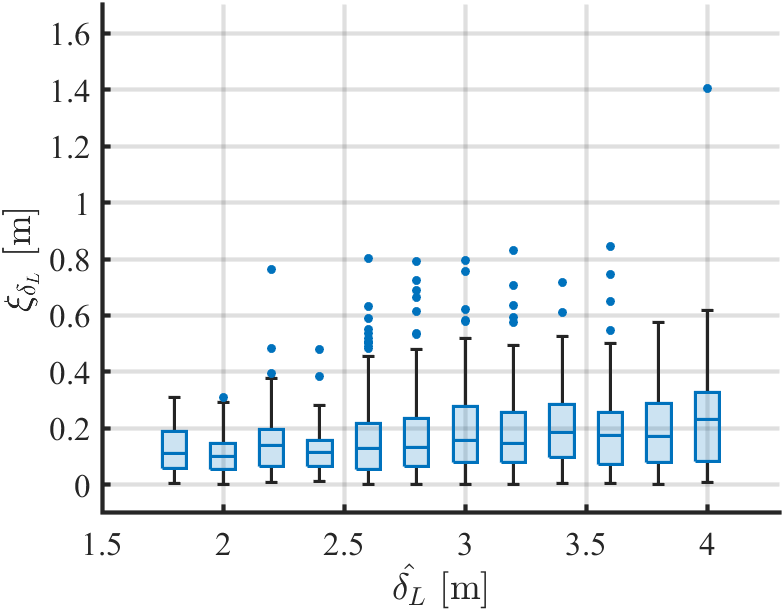}\label{fig:dl}}
    \caption{\em Absolute error of estimated distances $\delta_R$ and $\delta_L$ against the actual distances (a)$\hat{\delta_R}$ and (b)$\hat{\delta_L}$.}
    \label{fig:noisyd}
\end{figure}

We modeled the uncertainty in the $RSS$ produced by 
receiving either more or less power than the original $RSS$ as a Gaussian random variable $\Phi_{\sigma}$ with zero mean and a standard 
deviation $\sigma$ as follows:
\begin{equation*}\label{Eq:freespaceNoise}
      RSS'(Rx) = RSS(Rx) + \Phi_{\sigma},
\end{equation*}
where $RSS'$ is the noisy received signal strength at a generic transceiver $Rx$.
We set $\sigma$ to 0.3 mW since we only consider short transmission ranges and near-the-ground communication. The added noise affects the distance we 
estimate based on the $RSS$. We measured the impact of noise by calculating the absolute error $\xi$ of the estimated distance at each transceiver $RSS(R)$ 
and $RSS(L)$ where $\xi_{\delta_R} = |\hat{\delta_R} - \delta_R|$ and $\xi_{\delta_L} = |\hat{\delta_L} - \delta_L|$.
Fig.~\ref{fig:noisyd} shows the distribution of the absolute errors in distance estimated from $RSS'(R)$  and $RSS'(L)$ plotted against the actual 
distances (a) $\hat{\delta_R}$ and (a) $\hat{\delta_L}$ obtained from SUMO respectively. Note that the errors shown are observed at every 0.2-meters window. 
Results showed that the added noise causes minor distance estimation errors in ranges near the transceivers, but the errors increase as the pedestrians move 
away from the transceivers.

\begin{figure}[ht]
\centering
    \subfigure[\em The noise produces multiple $const$ values for the same parallel line to the sidewalk  while the actual $const$ is unique for each 
line.]{\includegraphics[width=0.3\textwidth]{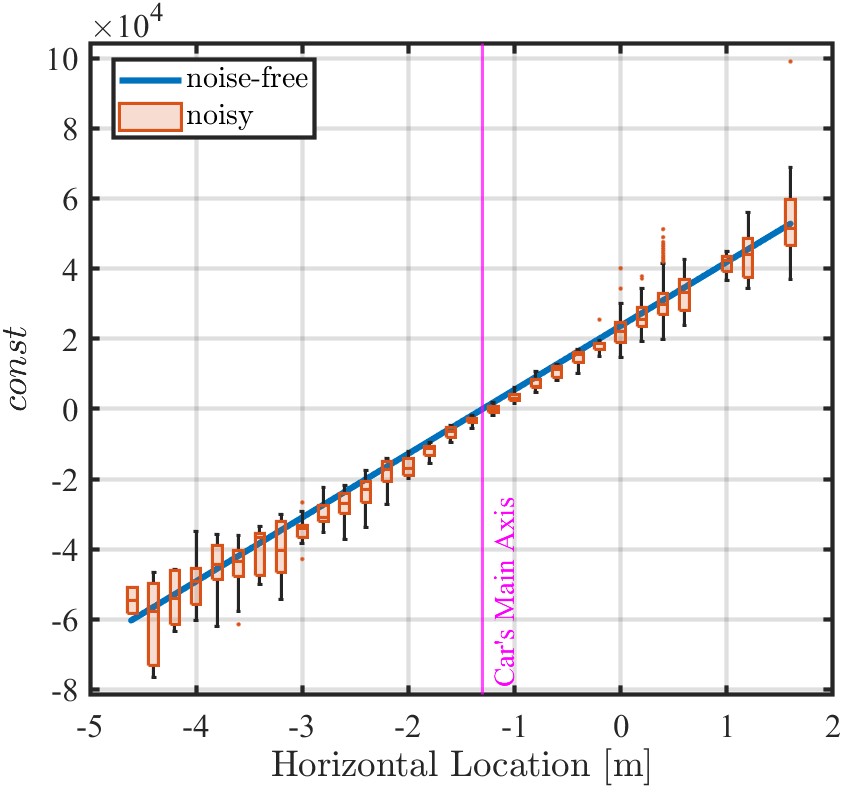}\label{fig:noisyconst}}
    \hspace{3em}
    \subfigure[\em Distribution of $const$ of pedestrians' signals per pedestrian's class (on the sidewalk and in the street). $c_0$ is used as a threshold to 
	classify pedestrians.]{\includegraphics[width=0.3\textwidth]{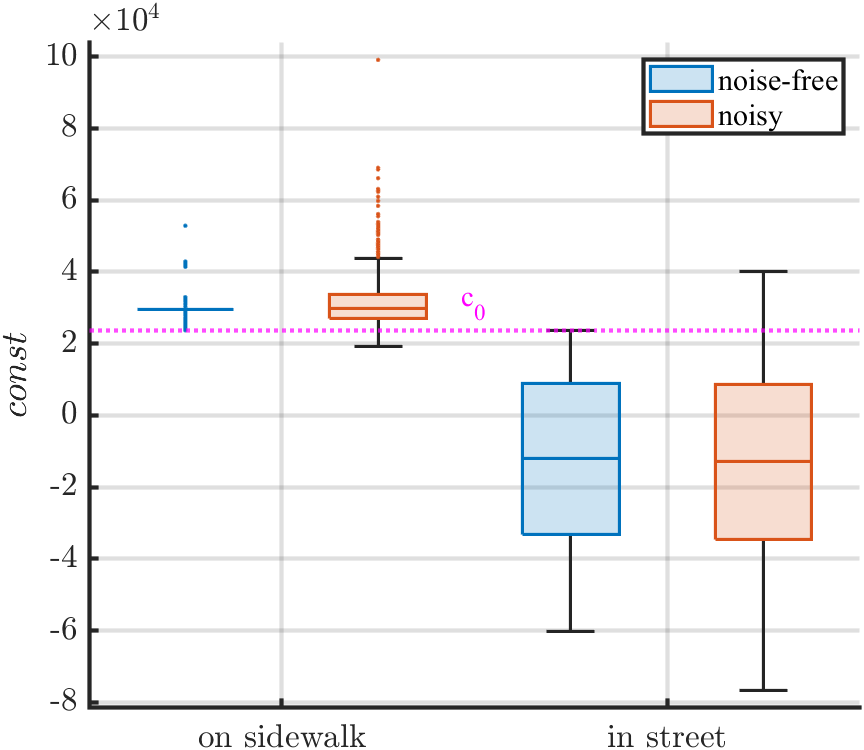}\label{fig:c0}}
 \caption{\em Effect of noise on $const$}
\end{figure}

As the noise affects the $RSS$ at both transceivers, it also affects the result of \eqref{RSS-1}. Consequently, the noise impacts the one-to-one mapping, 
discussed in Section~\ref{sec:class+loc}, between the set of lines parallel to the edge of the sidewalk
and the $const$ values as each line will have many $const$ values. To determine the range of inaccuracy in $const$, we define a 0.2-meters window of the 
horizontal position before and after the sidewalk and aggregate the calculated noisy $const$ from $RSS'(L)$ and $RSS'(R)$ at each window as we show in
Fig.~\ref{fig:noisyconst}. To compare the noisy $const$ with the noise-free $const$, we show in the same figure the noise-free $const$. 
The results showed that the noise impacts the $const$ if pedestrians are far from the main axis of the car (i.e. middle of the LR line segment), 
while it has a lower impact near the LR midpoint the pedestrians are close to both transceivers and $const$ is almost zero. This affected also the classification 
based on the threshold $c_0$. Referring to Fig.~\ref{fig:c0}, all $const$ values of pedestrians walking on sidewalk are greater than $c_0$. 
Similarly, all $const$ values of pedestrians who are in the street are less than $c_0$. However, with the noisy $const$, several $const$ values of pedestrians walking on the sidewalk 
are less than $c_0$, and several $const$ values of pedestrians walking in the street are greater than $c_0$.
%\begin{figure}[!h]
 %   \centering
%    \includegraphics[width=0.4\textwidth]{Figures/RecBoxPlot.png}
%    \caption{\em Distribution of $const$ of pedestrians' signals per pedestrian's class (on the sidewalk and in the street). $c_0$ is used as a threshold to 
%	classify pedestrians.}
 %   \label{fig:c0}
%\end{figure}

\section{Evaluation} \label{sec:evaluation}

We evaluated the performance of ADOPT in two scenarios: 
\begin{itemize}
\item {\bf Scenario 1 -- Noise-free mode:}  in which we performed pedestrian classification and crossing time estimation using  the actual data generated by SUMO;
\item {\bf Scenario 2 -- Noisy mode:} in which we performed pedestrian classification and crossing time estimation using noisy RSS to have a more realistic evaluation. 
\end{itemize}

We demonstrate the performance of ADOPT in both scenarios above. Specifically, in Subsection \ref{subsec:class} we discuss the 
accuracy of pedestrian classification; in Subsection \ref{subsec:local} we discuss the accuracy of pedestrian localization, once they are in the street;
the accuracy of pedestrian street traversal speed is discussed in Subsection \ref{subsec:speed}; the accuracy of the remaining crossing time
is discussed in Subsection \ref{subsec:remaning}; the accuracy of establishing the Safety Zone is discussed in Subsection \ref{subsec:safety-zone}. 
Finally, Subsection \ref{subsec:end-to-end} offers an end-to-end evaluation of ADOPT.

\subsection{Accuracy of pedestrian classification}\label{subsec:class}

We evaluated the accuracy of pedestrian classification as  {\em on the sidewalk} or {\em in the street} by using the following formula
\begin{equation}\label{eq:accuracy}
accuracy = \frac{TP+TN}{TP+TN+FP+FN} \times 100,
\end{equation}
where 
\begin{itemize}
\item TP (True Positive): is the total number of pedestrians that were correctly identified to be in the street;
\item TN (True Negative): is the total number of pedestrians that were correctly identified to be on the sidewalk;
\item FP (False Positive): is the total number of pedestrians incorrectly classified as in the street; and,
\item FN (False Negative): is the total number of pedestrians incorrectly identified as on the sidewalk.
\end{itemize}

\noindent
We obtained the ground truth of pedestrians classes directly from SUMO generated data.
Referring to Table~\ref{tab:overallaccuray} the accuracy of pedestrian classification in the noise-free mode is 100\% accuracy, while the accuracy dropped to 93.25\% in the noisy mode. 

%\begin{table}[h]
%\caption{Accuracy of Pedestrian Classification}
%\begin{center}
%\begin{tabular}{|l|c|}
%\hline
%\textbf{Scenario}       & \textbf{Classification Accuracy} \\
%\hline
%noise-free              & 100\%      \\
%\hline
%noisy                   & 92\%      \\
%\hline
%%\multicolumn{2}{1}{$^{\mathrm{1}}$ designated to detect pedestrians.}
%\end{tabular}
%\label{tab:overallaccuray}
%\end{center}
%\end{table}

\begin{table*}[]
\centering
\caption{Overall Accuracy of Pedestrian Classification}
\label{tab:overallaccuray}
\resizebox{\textwidth}{!}{%
\begin{tabular}{@{}llcccc@{}}
\toprule
\multicolumn{2}{l}{\multirow{2}{*}{}} &
  \multicolumn{2}{c}{\textbf{Predicted   Class (noise-free mode)}} &
  \multicolumn{2}{c}{\textbf{Predicted Class (noisy mode)}} \\ \cmidrule(l){3-4} \cmidrule(l){5-6} 
\multicolumn{2}{l}{} &
  \textbf{in street (positive)} &
  \textbf{on sidewalk   (negative)} &  
  \textbf{in street (positive)} &
  \textbf{on sidewalk   (negative)} \\ \cmidrule(l){3-4} \cmidrule(l){5-6}
\multirow{2}{*}{\textbf{Actual Class}} & \textbf{in street (positive)}     & 340 (TP)         & 0 (FN)          & 316  (TP)      & 24   (FN)       \\ %\cmidrule(l){3-4} \cmidrule(l){5-6}
                                       & \textbf{on sidewalk   (negative)} & 0  (FP)         & 1216  (TN)      & 81  (FP)      & 1135  (TN)      \\ \cmidrule(l){1-2} \cmidrule(l){3-4} \cmidrule(l){5-6} 
\multicolumn{2}{c}{\textbf{Accuracy}}                                               & \multicolumn{2}{c}{100\%} & \multicolumn{2}{c}{93.25\%} \\ \hline
\end{tabular}%
}
\end{table*}

%\begin{figure}[t]
%\centering
%    \subfigure[]{\includegraphics[width=0.23\textwidth]{Figures/NoisyConsts2.png}\label{fig:noisyconst}}
%    \hspace{0.2em}
%    \subfigure[]{\includegraphics[width=0.23\textwidth]{Figures/constacc1.png}\label{fig:classacc1}}
%    \caption{Impact of noise on (a)$const$ and the (b)classification performance}
%    \label{fig:noiseClass}
%\end{figure}

\begin{figure}[h]
    \centering
    \includegraphics[width=0.3\textwidth]{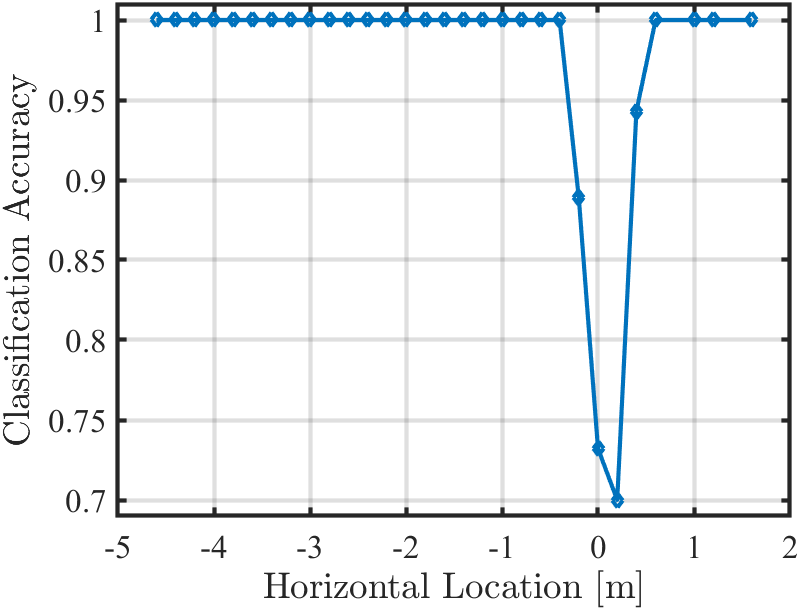}
    \caption{\em Aggregated classification accuracy at horizontal locations before and after the edge of the sidewalk.}
    \label{fig:classacc1}
\end{figure}

To investigate the impact of noise on the classification performance in noisy mode, in Fig.~\ref{fig:classacc1} we plot the classification accuracy against 
the horizontal location of the transmitted signal from the edge of the sidewalk that is indicated with value 0. Locations with negative values are in the street while 
locations with positive values are on the sidewalk. We noticed that the classification accuracy only drops when the transmitted signal is within few decimeters
away from the edge of the sidewalk (i.e. location 0).

\begin{figure*}[ht]
  \centering
  \subfigure[\em Detailed classification accuracy based on $const$ and $const'$]{\includegraphics[width=0.31\textwidth]{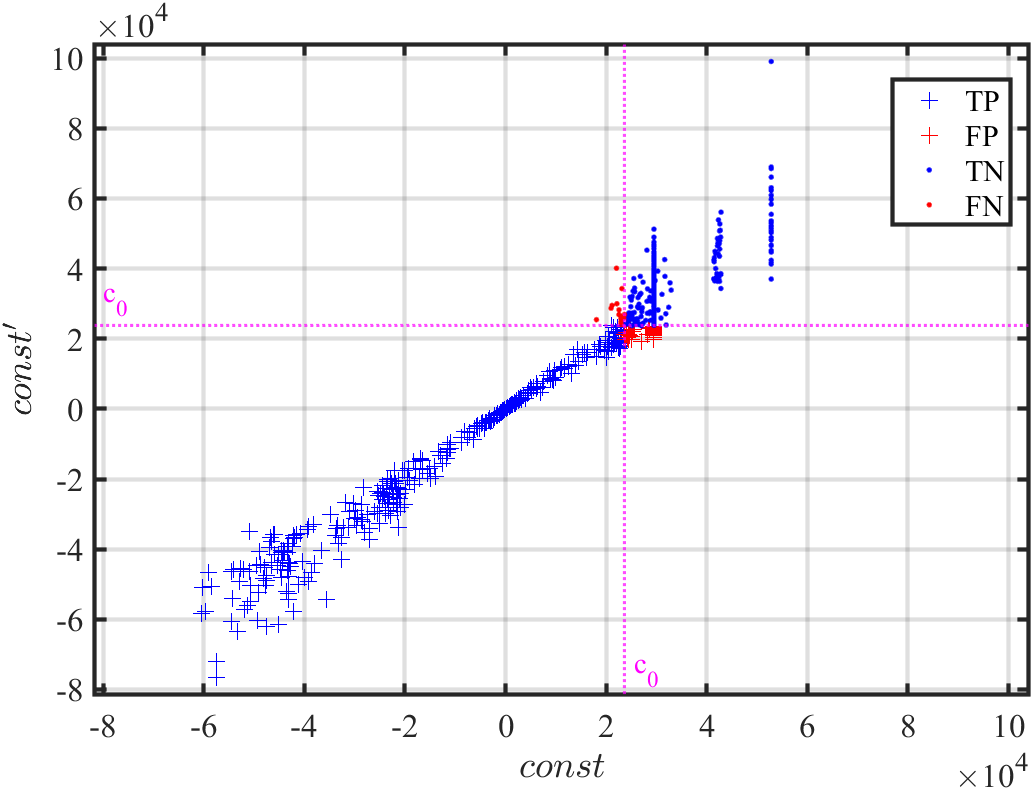}\label{fig:classacc}}
  \hspace{1em}
  \subfigure[\em Detailed classification accuracy at horizontal locations before and after the edge of the sidewalk]{\includegraphics[width=0.285\textwidth]{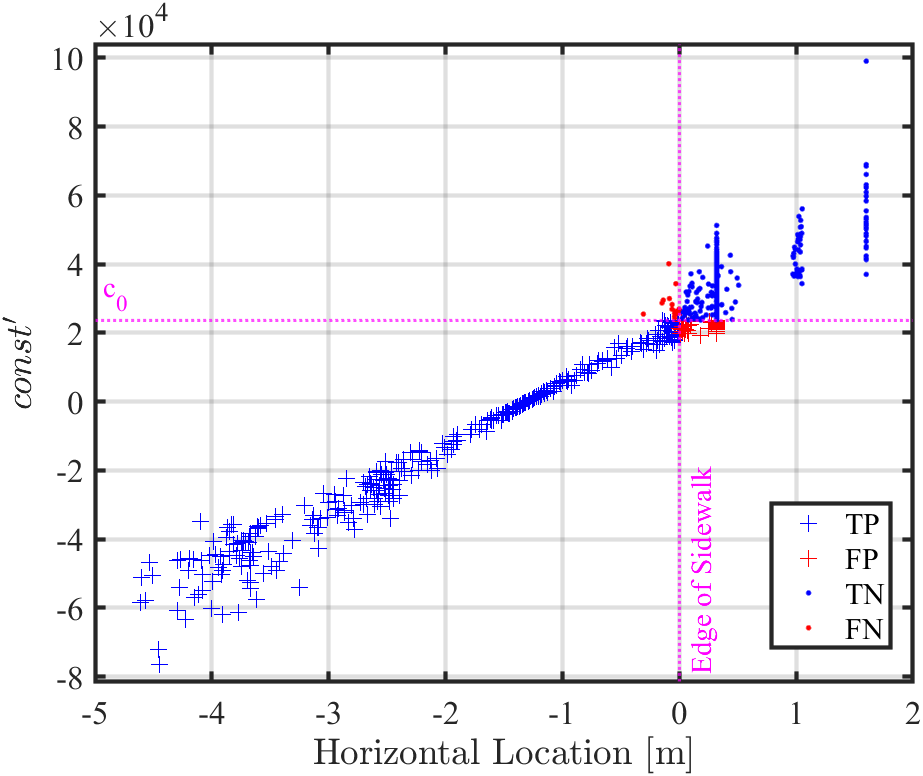}\label{fig:classaccy}}
  \hspace{1em}
  \subfigure[\em FPPL and FNPL are high only within 1 meter before and after the edge of the sidewalk with noisy signals.]{\includegraphics[width=0.285\textwidth]{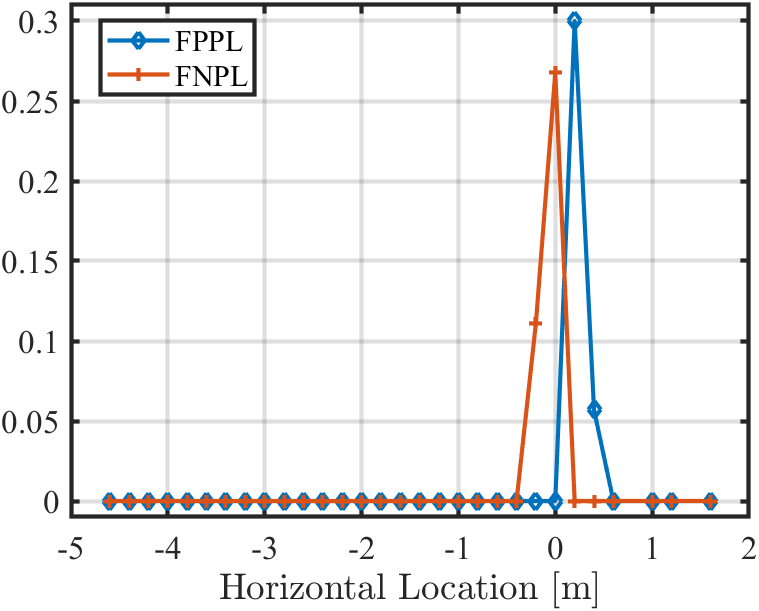}\label{fig:fnry}}
  \caption{\em ADOPT performance in pedestrian classification}
\end{figure*}

To better understand what is going on, we investigated the accuracy metrics TP, FP, TN, FN, defined above in more detail. 
Referring to Fig.~\ref{fig:classacc}, the accuracy drops when we have FP and FN due to the noisy $const$ (denoted by $const'$). We noticed that FP classification 
occurs if $const$ is less than $c_0$ and, at the same time, $const'$ is greater 
than $c_0$. This means we have FPs if the noise generates $const$ above the threshold $c_0$ while the actual $const$ is lower than the threshold. Similarly,  
FN classifications occur  if $const$ is greater than $c_0$ and, at the same time, $const'$ is less than $c_0$. This means we have FN if the noise generated $const'$ 
below the threshold $c_0$ while the actual $const$ is above the threshold. From the figure, we can see that this happens only in a limited area around $c_0$. 
To translate this to spatial data, in  Fig.~\ref{fig:classaccy} we plotted $const'$ corresponding to the horizontal location of the transmitted signal. 
%From this figure, we can see that the false classification happens only around 1 meter from the edge of the sidewalk. \par

%Fig.~\ref{fig:classacc} shows the accuracy metrics distributed among $const$ and $const'$ where $const'$ is the noisy version of $const$. The figure also show 
%the $c0$ value in dotted line. As it can be seen in the figure, when the actual $const$ is less than $c0$ and $const'$ is greater than $c0$, 
%we have FN for all the signals which decreases the accuracy. This explains why the accuracy based on the $const'$ is not always low when $const'$ is 
%greater than $c0$. We have also FP for all signals when the $const'$ is less than $c0$ and at the same time, $const$ is greater than $c0$. 
%To translate this into spatial data, we plot $const'$ corresponding to the lateral position of the transmitted signal in Fig.~\ref{fig:classaccy}. 
%From this figure, we can see that the false classification happens only around a few decimeters from the edge of the sidewalk. \par

Obviously, high FN puts pedestrians at the risk as they enter the street and ADOPT does not alert the approaching cars. On the other hand, high FP results 
in flooding  the approaching vehicles with incorrect alert messages while the pedestrians are on the sidewalk. To assess where ADOPT has 
miss-classification, we define our metric False Positive Per Location FPPL and False Negative Per Location FNPL where FPPL is the percentage of FP per horizontal 
location aggregated at each 0.2 meters, and FNPL is the percentage of FN per horizontal location aggregated at each 0.2 meters as well. Fig.~\ref{fig:fnry} shows that 
FNPL is high only at locations situated a few decimeters away from the edge of the sidewalk. 
Moreover, ADOPT has higher FPPL a few decimeters away from the edge of the sidewalk. % based on the amount of noise we added. 
In conclusion, the ADOPT pedestrian classification scheme is able to classify pedestrians accurately even in the presence of  noisy signals. 
%The possibility of incorrect classification is restricted to positions 1 meter away from the edge of the sidewalk.

The other metrics we use to evaluate empirically the performance of  ADOPT are: the accuracy of pedestrian localization, the accuracy of crossing time estimation, 
the accuracy of the remaining crossing time, as well as the accuracy of the Safety Zone size. 
We use Root Mean Squared Error (RMSE) to measure the accuracy of these estimations. We show the overall result of the RMSE in Table~\ref{tab:time-to-cross eval}. 
In the following subsections, we discuss each of them in detail.

%\begin{table}[t]
%\caption{\em ADOPT Estimation RMSE}
%\begin{center}
%\begin{tabular}{|c|c|c|c|c|c|}
%\hline
%\textbf{Mode}                         & \textbf{$E_y$  [m]} & \textbf{$E_d$  [m]}& \textbf{$E_v$ [m/sec]} & \textbf{$E_\Delta$[sec]} & \textbf{$E_D$[m]}\\
%\hline
%NF               &       0.00 &       0.00 & 0.03 & 0.49 & 10.68\\
%\hline
%N                     &      0.28 & 0.17 & 0.06 &  0.96 & 17.59\\
%\hline
%\multicolumn{2}{1}{$^{\mathrm{1}}$ designated to detect pedestrians.}
%\end{tabular}
%\label{tab:time-to-cross eval}
%\end{center}
%\end{table}

\begin{table}[t]
\caption{\em ADOPT Estimation RMSE}
\begin{center}
\begin{tabular}{ccc}
\hline
\textbf{RMSE} & \textbf{noise-free mode} &  \textbf{noisy  mode}\\ \hline
$E_y$ & 0.00 m &  0.26 m \\
$E_d$ & 0.00 m & 0.24 m\\
$E_v$ & 0.05 m/sec &0.11 m/sec \\
$E_\Delta$ & 0.51 sec &  1.21 sec\\
$E_D$ &   12.79 m & 42.23 m\\
\hline
%\multicolumn{2}{1}{$^{\mathrm{1}}$ designated to detect pedestrians.}
\end{tabular}
\label{tab:time-to-cross eval}
\end{center}
\end{table}

\subsection{Accuracy of pedestrian localization}\label{subsec:local}

To measure the overall accuracy of pedestrian localization, we evaluated $E_d$ and $E_y$, where $E_d$ is the RMSE of location of pedestrians in front of 
the parked car (the vertical distance of the pedestrian) and $E_y$ is the RMSE in estimating the pedestrian's 
distance from the edge of the sidewalk (the horizontal distance of crossing pedestrians) as follows:

We evaluated $E_{d}$ as follows: 
$$
E_{d} = \sqrt \frac{\Sigma^M_{j=1} (\hat{d}_j-d_j)^2}{M}
$$ 
where  $M$ is the total number of received signals and $\hat{d}_j$ and $d_j$ are, respectively, the actual and estimated vertical distances of the received signal. \par
Similarly, we calculate: 
$$
E_{y} = \sqrt \frac{\Sigma^{M_s}_{j=1} (\hat{y}_j - y_j)^2}{M_s}
$$
where $M_s$ is the total number of received signals that are actually detected in the street.  $\hat{y}_j$ and $y_j$ are the actual and estimated horizontal distances. Recall 
that $y$ is only calculated when a pedestrian is classified as in the street. 

\begin{figure}[ht]
\centering
    \subfigure[\em $\xi_d$ against $\hat{d}$]{\includegraphics[width=0.5\textwidth]{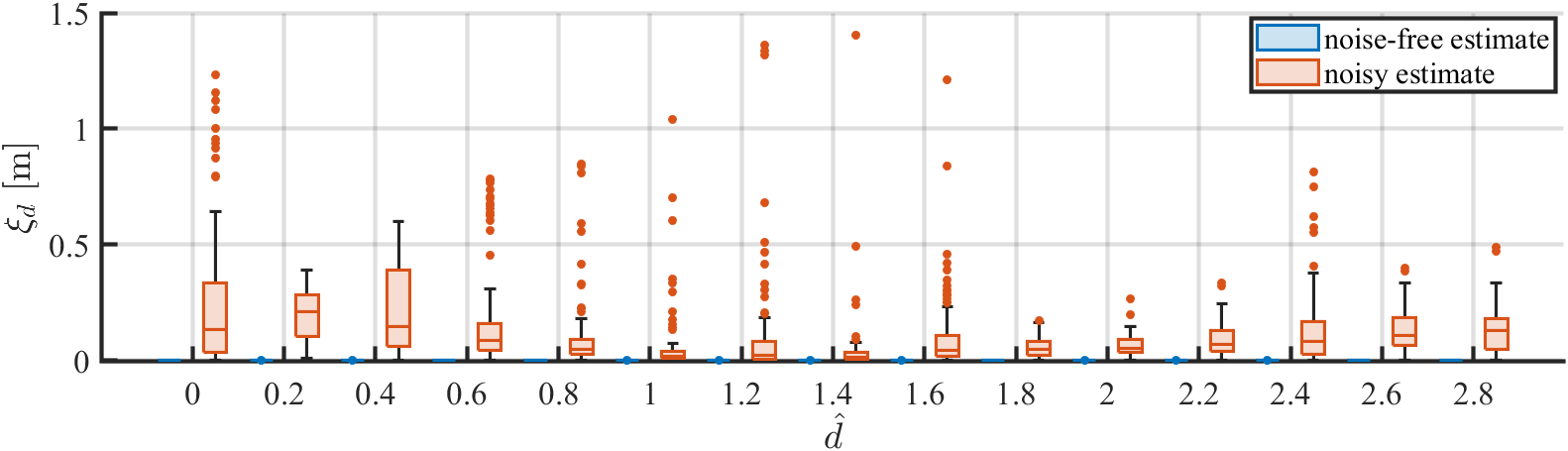}\label{fig:dcoorderr}}
    \subfigure[\em $\xi_y$ against $\hat{y}$]{\includegraphics[width=0.5\textwidth]{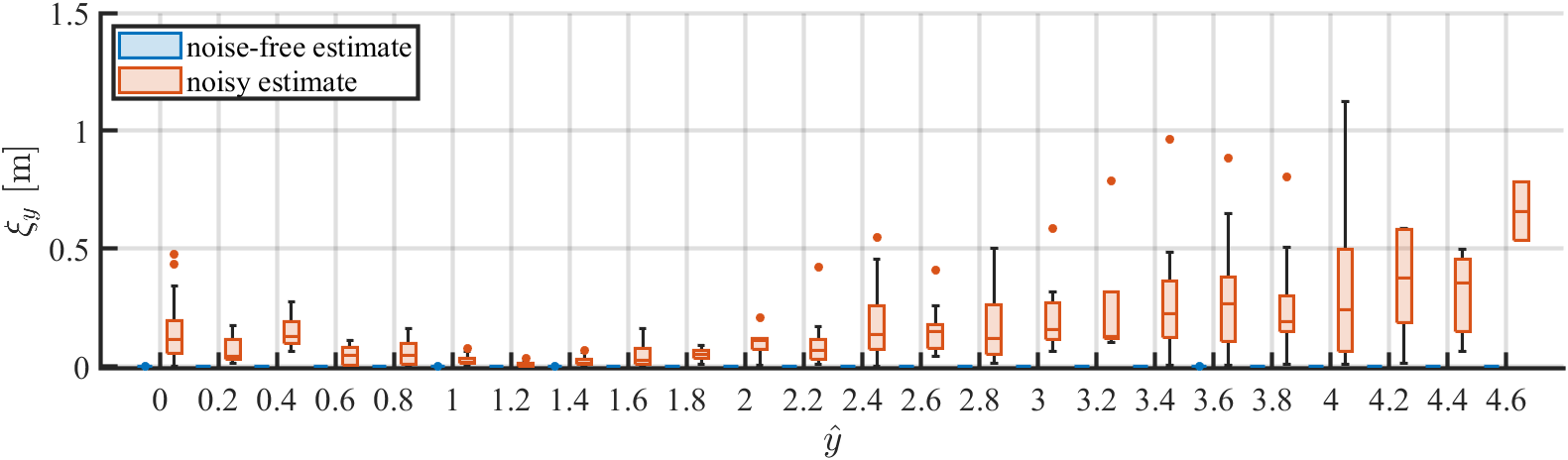}\label{fig:ycoorderr}}
    \caption{\em Distribution of localization error against the actual location}
\end{figure}

To show the result in detail, we determined a 0.2-meters window of $\hat{d}$. 
Fig.~\ref{fig:dcoorderr} shows that the aggregated absolute error 
$
\xi_d = |\hat{d} - d|
$ 
is zero in the noise-free mode and does not exceeds 1.5 meters in the noisy mode. 

Similarly, we determined a 0.2-meters window of $\hat{y}$ to observe the absolute error at each window. The absolute error of $y$ is 
$
\xi_y = |\hat{y} - y|
$. 
Fig.~\ref{fig:ycoorderr} shows that the estimation of $y$ in the noise-free mode is accurate since we have zero aggregated error. However, the aggregated error increases in noisy  mode. The errors are lower as pedestrians start crossing the street because they are closer to the main axis of the car (i.e. around $y=1.3$) and the noise is low in this area as we showed previously in Fig.~\ref{fig:noisyconst}. 

\subsection{Accuracy of crossing  speed estimation}\label{subsec:speed}

\begin{figure}[ht]
\centering
    \subfigure[\em Dynamic Cohort Size]{\includegraphics[width=0.3\textwidth]{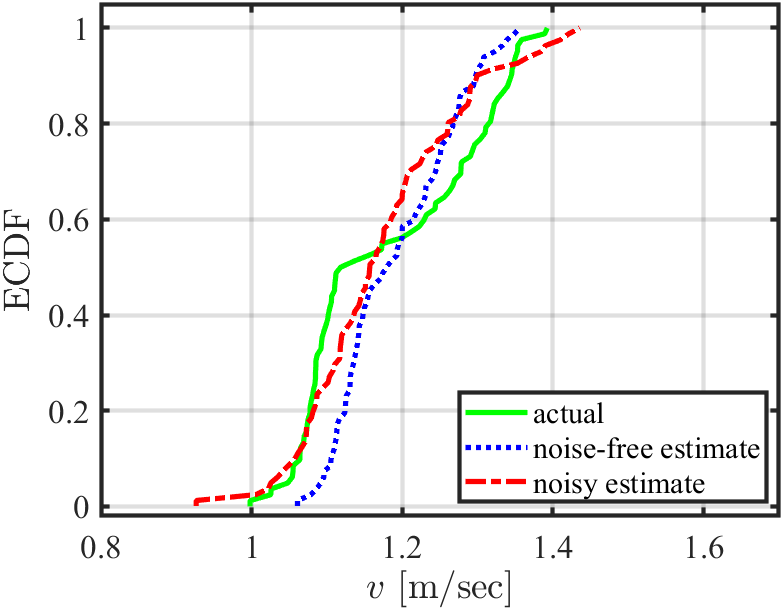}\label{fig:speeds}}
    \hspace{2em}
    \subfigure[\em Fixed Cohort Size]{\includegraphics[width=0.3\textwidth]{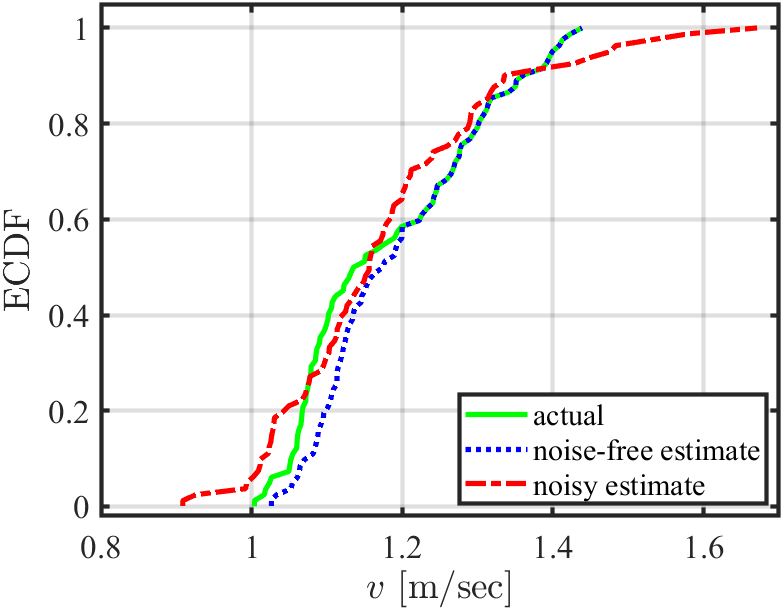}\label{fig:speedso}}
    \caption{\em ECDF of Pedestrian Speeds}
\end{figure}

We used RMSE to measure the overall accuracy of estimating the pedestrian's speed $v$ as follows:
$$
E_{v} = \sqrt \frac{\Sigma^N_{i=1}( \hat{v}_i-v_i)^2}{N}
$$
where $N$ is the total number of pedestrians generated in the simulation, and $\hat{v}_i$ and $v_i$ are, respectively, the actual and the estimated speed of each pedestrian 
who was actually in the street. 
We chose to average the speed of each pedestrian individually because one pedestrian may vary her speed in SUMO.
Fig.~\ref{fig:speeds} shows the Empirical Distribution Cumulative Function ECDF of the actual speeds, that are retrieved from SUMO, and the speeds that we estimate in noise-free and lastly the speed estimate in the noisy mode. The results showed that the noise affects the speed estimation as we can see from the difference between 
the actual and the noisy estimates in the figure. In addition, we noticed that the difference between the actual and noise-free estimates is due to the 
change of the crossing cohort size, and this happens only when a new pedestrian joins the cohort. We show in Fig.~\ref{fig:speedso} 
the ECDF after removing the samples where a new pedestrian joins the cohort. As it can be seen in the figure, the difference between the noise-free 
estimated speed and the actual speed is lower when the cohort size is fixed.

\subsection{Accuracy of the remaining crossing time}\label{subsec:remaning}
\begin{figure}[ht]
    \centering
    \subfigure[\em Dynamic cohort size]{\includegraphics[width=0.3\textwidth]{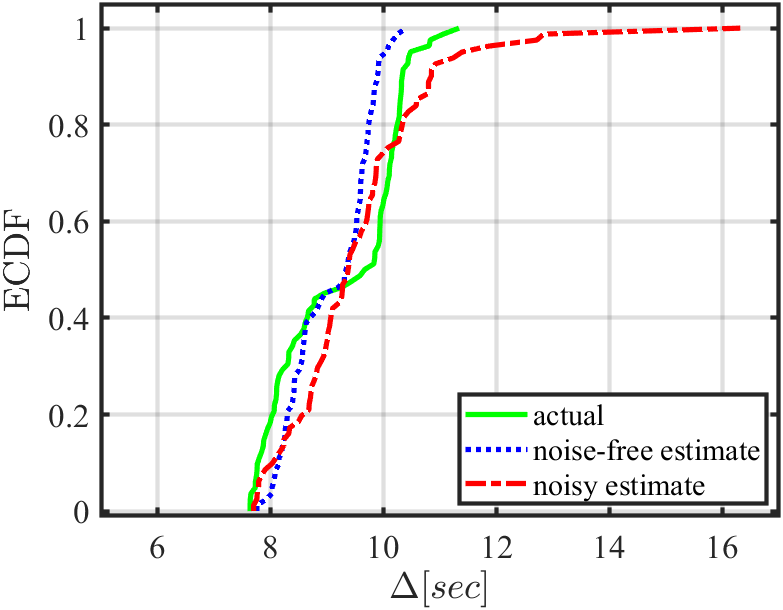}\label{fig:tcs}}
    \hspace{2em}
    \subfigure[\em Fixed cohort size]{\includegraphics[width=0.3\textwidth]{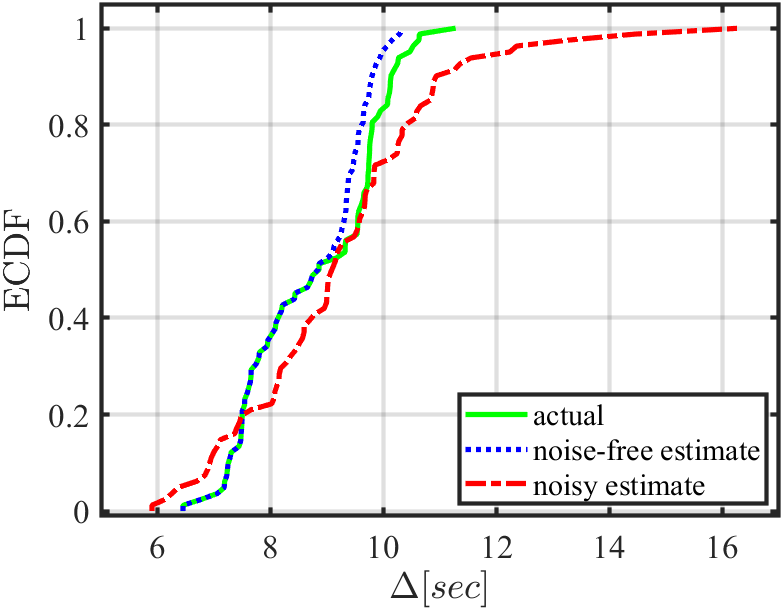}\label{fig:tcso}}
    \caption{\em ECDF of remaining crossing time}
\end{figure}

We used the same RMSE formula to measure the accuracy of the remaining crossing time:
$$
E_{\Delta} = \sqrt \frac{\Sigma^N_{i=1} (\hat{\Delta}_i-\Delta_i)^2}{N}
$$
where $\hat{\Delta}_i$ and $\Delta_i$ are, respectively, the actual and estimated remaining time to cross for each pedestrian.
Fig.~\ref{fig:tcs} shows the ECDF of the average crossing time for each pedestrian in the street. We calculated the actual remaining crossing time based on the 
actual speed retrieved from SUMO. As the noise affects the speed, it also affects the remaining crossing time as we show in the figure. We noticed also 
that the difference between the actual and noise-free estimation is due to the change of the crossing cohort size, and this happens only the first time a new 
signal is detected in the crossing cohort. We show in Fig.~\ref{fig:tcso} the ECDF after removing the samples where a new pedestrian joins the cohort. 
As it can be seen from the figure, the difference between the noise-free estimate and the actual time to cross is lower.

\subsection{Accuracy of Safety Zone size}\label{subsec:safety-zone}

The RMSE of estimating the Safety Zone size, which is determined by the propagation distance $D(t)$, for each pedestrian is calculated as follows: 
$$
E_{D} = \sqrt \frac{\Sigma^N_{i=1} (\hat{D}_i-D_i)^2}{N}
$$
where $\hat{D}_i$ and $D_i$ are, respectively, the actual and estimated propagation distances. $\hat{D}_i$ is calculated based on the 
actual remaining time to cross $\hat{\Delta}_i$. 
\begin{figure}[h]
    \centering
    \subfigure[\em Dynamic cohort size]{\includegraphics[width=0.3\textwidth]{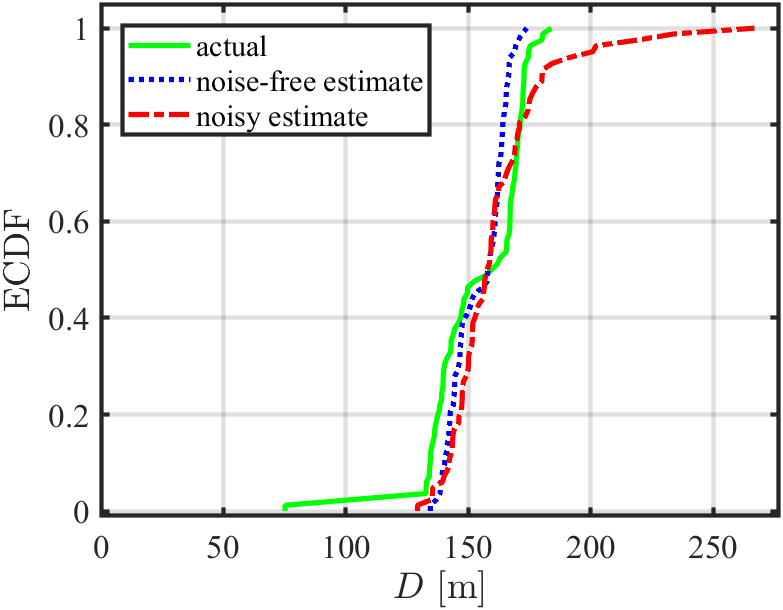}\label{fig:pd}}
    \hspace{2em}
    \subfigure[\em Fixed cohort size]{\includegraphics[width=0.3\textwidth]{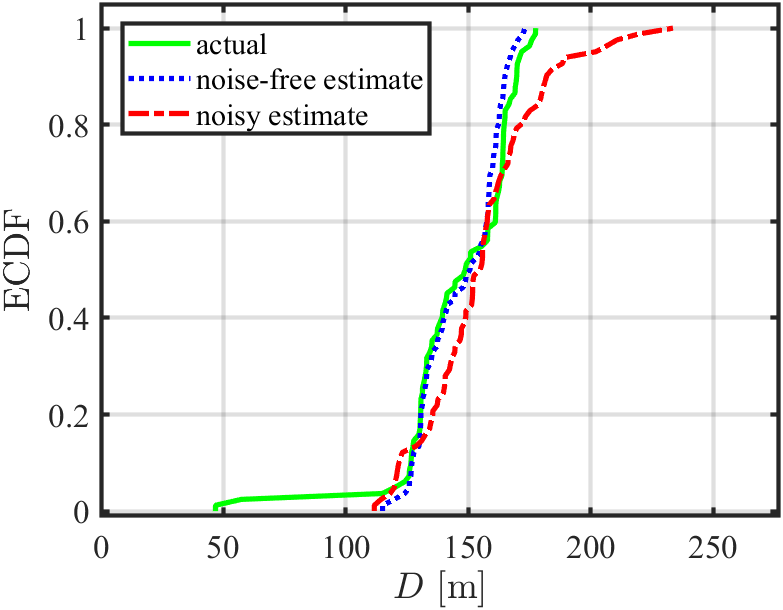}\label{fig:pdoneped}}
    \caption{\em ECDF of propagation distance.}
    \label{fig:pgdistecdf}
\end{figure}
Fig.~\ref{fig:pgdistecdf} shows the ECDF for the propagation distance estimation averaged for each pedestrian in the dynamic and fixed cohort sizes. We noticed that the difference 
between the actual and the noise-free estimate is less in the fixed cohort size. The difference between them increases based on the error of 
$\Delta$ estimation. \par
\begin{figure}[ht]
    \centering
    \includegraphics[width=0.5\textwidth]{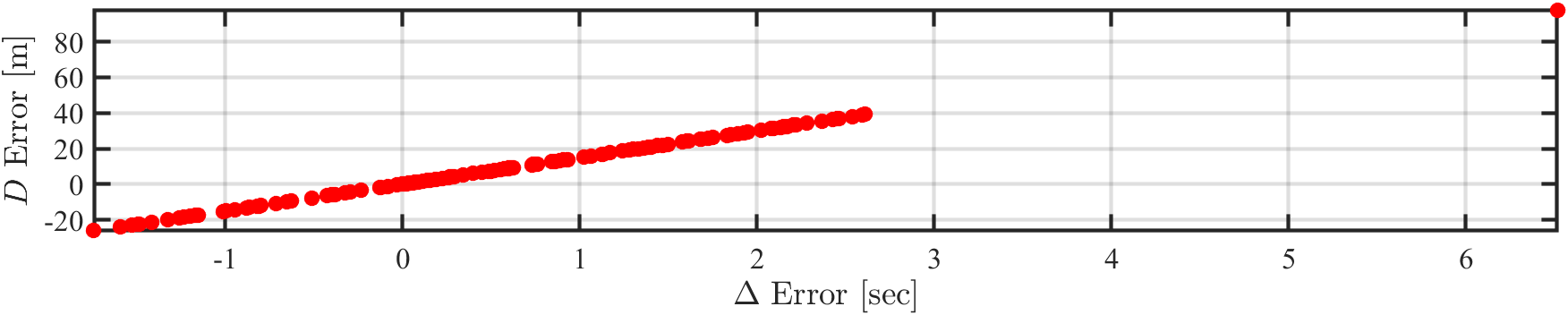}
    \caption{\em Correlation between the estimation error of $D$ and $\Delta$ }
    \label{fig:relErrorpdtc}
\end{figure}

To justify the difference between the actual and estimated propagation distance, we measured the relative error of estimating 
$\Delta$ and $D$ in noise-free mode for fixed size cohort compared to their actual values where the $relative~error = actual~value - estimated~value $.
%We chose noise-free because we omit the effect of noise in our reasoning, and we chose fixed cohort size to omit the effect of a new pedestrian 
%joining the cohort. 
The results showed that there is a high 
correlation between the relative error of estimating $\Delta$ and $D$ as we show in Fig.~\ref{fig:relErrorpdtc}. The figure also shows 
that the majority of $\Delta$ errors are less than 3 seconds and cause about additional 40 meters of $D$. We see a high error in $D$ since each 
$\pm 1$ second of error in $\Delta$ is multiplied by the car's speed and the added $r$, so this second produces about $\pm 15$ meters of $D$ error. If the error is positive then ADOPT estimates a larger safety zone and the pedestrian is safe. Indeed, the risk may increase if the estimation is less than needed.

In conclusion, the ADOPT classification accuracy depends on the amount of RSS noise that affects the $const$ values near the edge of the sidewalk. 
For localization, the accuracy of ADOPT depends on the RSS noise and the distance of the pedestrian from the transceiver receiving the strongest signals. 
ADOPT accuracy in estimating the speed, remaining crossing time and Safety Zone size depends on the noise and changes in cohort size.

\subsection{Evaluation of occluded pedestrian protection}\label{subsec:end-to-end}

To evaluate the performance of ADOPT as an end-to-end system to protect occluded crossing pedestrians, we calculated the speed of the approaching car upon 
reaching the pedestrian crossing area. We assumed that the driver or the automated vehicle react to the "Caution" message promptly  upon receiving it from the ADOPT
app running in the car. Note that SUMO has implemented a collision avoidance model to force vehicles to reduce their speed if there are crossing pedestrians. 
We disabled this mode before we generated the vehicular traffic. % and the car keeps moving at its speed $v_{crusing}$ even though there are crossing pedestrians.

\begin{figure}[ht]
    \centering
    \includegraphics[width=0.5\textwidth]{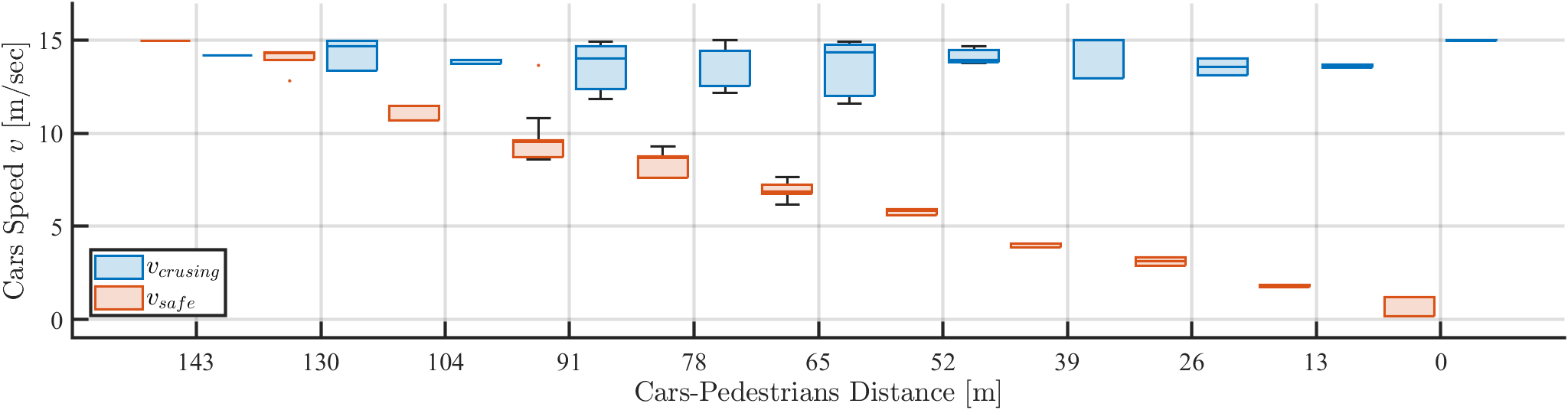}
    \caption{\em All approaching cars maintained safe speeds upon receiving the "Caution" message from ADOPT while they keep their cruising speeds without ADOPT "Caution" messages.}
    \label{fig:overallend2end}
\end{figure}

Referring to Fig.~\ref{fig:overallend2end}, the approaching cars generated in SUMO adopt the new speed $v_{safe}$ upon receiving the "Caution" messages as they approach 
the crossing pedestrians. We observed and aggregated the speeds at every 13-meters of car-to-pedestrians distances. As it can be seen also in the figure, the cars 
start maintaining their safe speeds gradually while they are approaching the crossing pedestrians which allows a smooth speed reduction without the need for a sudden stop. 
To compare the speed reduction caused by ADOPT with the speed of cars without ADOPT, we plotted their cruising speeds $v_{cruising}$ from SUMO in the same figure. 
We called it $v_{cruising}$ since SUMO cars do not maintain a fixed speed while moving, and also each car has its own speed 
random distribution with a determined maximum speed equal to the street's speed limit $v_{max}$. 

In the following, we show some examples from the simulation of approaching cars receiving ADOPT "Caution" message and changing their speed accordingly:
\begin{figure*}[ht]
 \centering
 \subfigure[\em One pedestrian crossing the street.]{\includegraphics[width=0.3\textwidth]{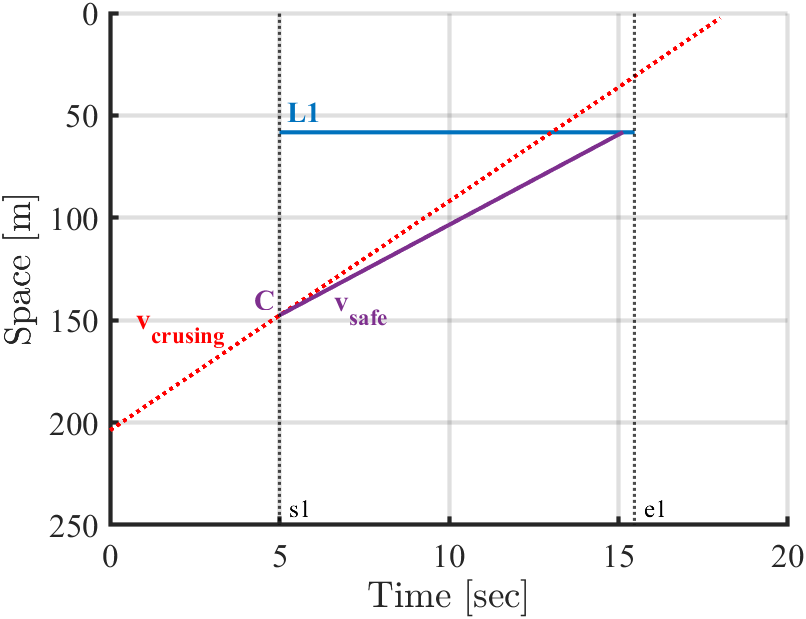}\label{fig:end to end 2}}
 \hspace{1em}
  \subfigure[\em Multiple pedestrians crossing the street at the same location]{\includegraphics[width=0.3\textwidth]{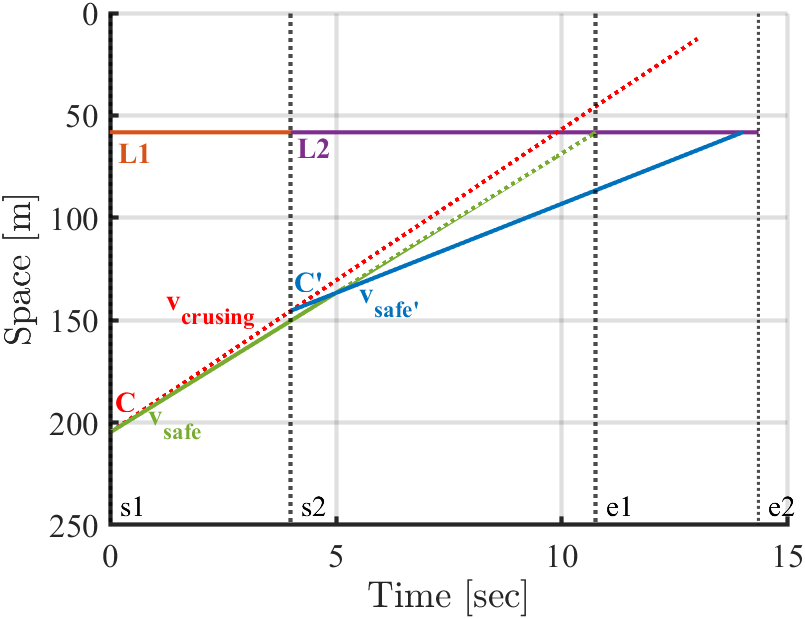}\label{fig:end to end 3}}
  \hspace{1em}
  \subfigure[\em Multiple pedestrians are crossing the street at different locations.]{\includegraphics[width=0.3\textwidth]{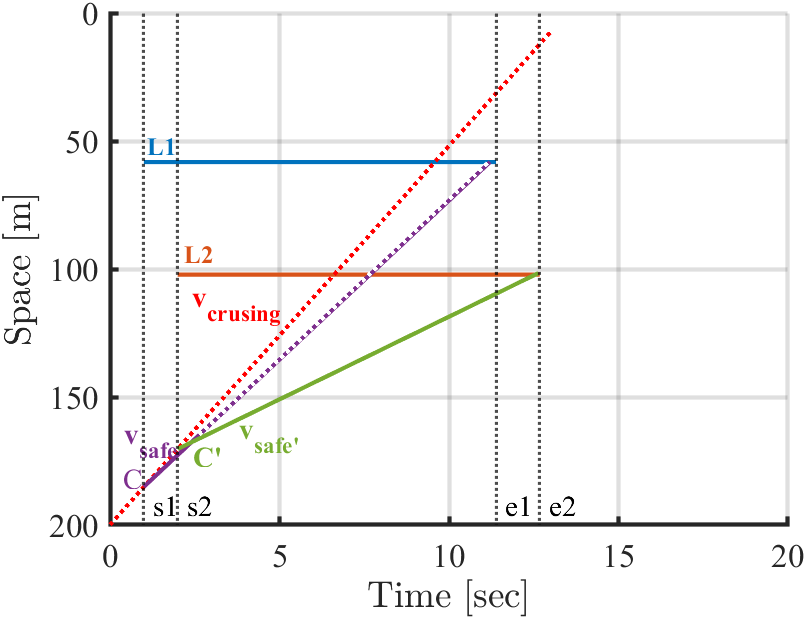}\label{fig:end to end 4}}
  \caption{\em Moving vehicle reduces its speed upon receiving "Caution" message from ADOPT in many cases.}
  \label{fig:end2end example1}
\end{figure*} 

In Fig.~\ref{fig:end to end 2}, the approaching car with speed $v_{max}$ receives a ``Caution'' message at location $C$ and time $s1$ for a pedestrian crossing at 
location $L1$. The car reduces its speed to $v_{safe}$ as a response to the alert message. We noticed that the car may intersect slightly before $(L1,e1)$ due 
to the error in estimating $d$ and the crossing time $\Delta$. In another example 
Fig.~\ref{fig:end to end 3}, the car receives another "Caution" message at location $C'$ as a new pedestrian joins the cohort from the same location of $L$, 
and finishes crossing at $e2>e1$. In this case, the car finds that the new speed (denoted by $v_{safe'}$) is lower than $v_{safe}$, so it adopts the new speed 
$v_{safe'}$ to avoid the collision. The expected reaction of the approaching car is to reduce its speed because the crossing time now becomes longer as a new 
pedestrian joins the cohort.

In a more complex example, cars report pedestrians at different locations $L1$ and $L2$ as in Fig.~\ref{fig:end to end 4}.
When the car receives a new ``Caution'' message  at time $s2$ and location $C'$, it calculates the new speed and finds that it is less than its 
current speed because $L2$ is closer than $L1$. Thus, it adopts the new speed $v_{safe'}$ until the pedestrian finish crossing.

In conclusion, approaching cars were able to maintain low speeds to avoid collision with the occluded crossing pedestrian accurately and in a timely manner based 
on the ``Caution'' messages received from ADOPT. This proves that ADOPT works end-to-end to effectively protect crossing pedestrians.

\section{Concluding remarks and future work}\label{sec:concl}

The common philosophy of all the pedestrian detection approaches of which we are aware is that this task should be undertaken by the moving cars 
themselves. In sharp 
departure from this philosophy, we proposed to employ cars parked along the sidewalk to detect and protect crossing pedestrians. 

In support of this goal, we have proposed ADOPT: a system for Alerting Drivers to Occluded Pedestrian Traffic.
ADOPT lays the theoretical foundations of a system that uses the on-board resources of parked cars to: 
\begin{itemize}
\item Detect the presence of a group of crossing pedestrians -- a crossing cohort; 
\item Predict the time the last member of the cohort takes to cross the street; 
\item Send alert messages to those approaching cars that may reach the crossing area while pedestrians are still in the street;
\item Show how approaching cars can adjust their speed to avoid crashing into crossing pedestrians.
\end{itemize}
\vspace*{1mm}

\noindent
Importantly, in  ADOPT all communications occur over very short distances and at very low power.
Our extensive simulations using SUMO-generated pedestrian and car traffic, have shown the effectiveness of ADOPT in detecting and protecting 
crossing pedestrians.

In spite of this, there are a number of  topics that need more work. First, the problem of information overload has to be mitigated. The problem arises when the 
(human) driver of an approaching car is alerted to the presence of various crossing cohorts. We are planning to design an app that minimized the information
overload and, consequently, driver distraction.

Second, it is of interest to optimize the process of disseminating alert messages 
as a function of the residual crossing time. In the current
version of the paper ``All clear'' messages informing approaching cars that the cohort has finished crossing are not used. Incorporating them into ADOPT
is targeted for future work.

Third, security and privacy are very important and are getting active attention as discussed in the next subsection.

\subsection{Security and Privacy in ADOPT}\label{subsec:SandP}

ADOPT involves two types of wireless communications: V2P and V2V. In the following, we discuss the security aspects of our communication types.

\begin{itemize}

\item Pedestrian to parked vehicle communications:
In the pedestrian detection and localization process, the system reads transmitted signals and makes decisions based solely on the signal strength and not on the
identity of the pedestrians. These decisions are based on anonymous received signal strengths that do not require
unique identifiers. Also, ADOPT detects and localizes pedestrians in short-range communications that do not require transmitting the
pedestrians' private data. With the proposed mechanism of pedestrian localization, ADOPT preserves pedestrians' privacy and security.
In spite of this, we may consider a potential attack that may be mounted against ADOPT such as a Denial of Service (DoS) attack. In this attack, a
group of pedestrians may stomp their feet on the
ground to activate the system. Our approach to pedestrian classification can distinguish if the signal is coming from the sidewalk or the street.
If this group of people is generating signals from the street, they are at risk and the system should notify approaching cars regardless of their intent.

\item Parked vehicle to approaching vehicle communication:
This type of communication involves the known security threats in V2V communications \cite{Arif2019-ag}. However, the short-range communication used in our system should
allow the use of frequency hopping \cite{Olariu2004-ar} to prevent attackers from sniffing or injecting fake information into V2V network. Moreover, short-range
V2V communications have their security advantages \cite{Rawat2014-ha} that are not present in long-range V2V communications.
It is of great interest to develop security primitive that leverage the 
type of short-rage communications that are used throughout the system \cite{samy-2012,yan-2012,notice-samy-2015}. 

\end{itemize}

%While security solutions developed over the years by the vehicular networking community can be used as a first line of defense, many of these involve 
%communications over long distances, which runs against the spirit of ADOPT. It is of great interest to develop security primitive that leverage the 
%type of short-rage communications that are used throughout the system \cite{samy-2012,yan-2012,notice-samy-2015}. 

\bibliographystyle{IEEEtran}
\bibliography{References}

\begin{appendix}

\section{Appendices}\label{sec:appendices}

\begin{figure}[ht]
    \centering
    \includegraphics[width=0.3\textwidth]{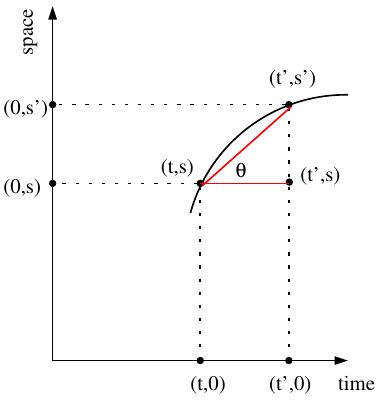}    
     \caption{\em Illustrating the time-space diagram.}
    \label{fig:ts-diagram}
\end{figure}

\subsection{The time-space diagram}\label{subsec:ts}

One of the basic tools in the tool-box of traffic engineers is the {\em time-space diagram} that allows one to plot the trajectories of vehicles
as curves in a Cartesian plane with axes labeled ``time'' and ``space''. Referring to Figure \ref{fig:ts-diagram}, the vehicle's coordinates
at time $t$ are $(t,s)$. The projection of this point on the vertical axis, $(0,s)$, indicates the position, at time $t$, of a vehicle that moves along the space
axis North-bound (i.e. from bottom to top). 

Assume that the same vehicle has continued moving along its trajectory in such a way that at time $t', (t'>t)$, its coordinates
in the time-space coordinate system are $(t',s')$. Equivalently, in the time interval $[t,t']$, the vehicle has moved along the space axis from location $(0,s)$ to
location $(0,s')$.

One of the nice features of the time-space diagram is that it allows one to compute and to {\em visualize} the average speed of the vehicle. Indeed, referring again
to  Figure \ref{fig:ts-diagram}, elementary physics indicate that the average speed, $v_{avg}$, of the vehicle in the time interval $[t,t']$ is the ratio
\begin{equation}\label{avg-speed}
v_{avg} = \frac{s'-s}{t'-t}.
\end{equation}

This is the same as the {\em slope} of the line segment connecting the points of coordinates $(t,s)$ and $(t',s')$. Equivalently, the average speed $v_{avg}$ 
is $\tan(\theta)$, where $\theta$
is the angle determined by the line segment connecting the points of coordinates $(t,s)$ and $(t',s')$ and the positive direction of the time axis.
\vspace*{2mm}

Finally, it is not hard to see that the {\em instantaneous speed} of the vehicle at an arbitrary time $\tau,\ (t \leq \tau \leq t')$ turns out to be the slope 
of the tangent to the trajectory at time $\tau$. In particular, if the trajectory happens to be a straight line, then the average speed matches the instantaneous speed, as 
expected.

\subsection{Time-space diagram of a crossing cohort}\label{subsec:cohort}

\begin{figure}[ht]
    \centering
    \includegraphics[width=0.3\textwidth]{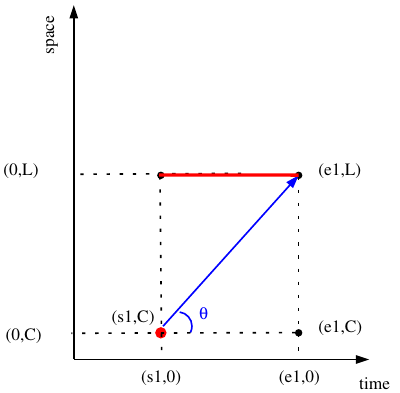}
     \caption{\em Illustrating the time-space diagram of a crossing pedestrian.}
    \label{fig:cp}
\end{figure}

Let us turn our attention to the time-space diagram corresponding to a crossing cohort. Referring to Figure \ref{fig:cp}, imagine a crossing cohort at location 
$L$ that starts crossing the street at time $s1$ and clears the street by time $e1$. Since the ``space'' coordinate of the cohort does not change, the corresponding 
time-space diagram is captured by a horizontal line segment with endpoints $(s1,L)$ and $(e1,L)$. 

Now, consider an approaching car and assume that the coordinates of the car when it receives the ``Caution -- pedestrian in the street'' message are $(s1,C)$.
What is the largest average speed that the approaching car should adopt to avoid crashing into the cohort?  The answer is simple: the car should not reach the point
$(0,L)$ while the crossing is in progress, namely in the time interval $[s1,e1)$. However, the car may reach $(0,L)$ at time $e1$,  as by that time the cohort has finished
crossing safely. Thus, using (\ref{avg-speed}),
the car should adopt the following {\em safe  average speed}:
\begin{equation}\label{safe-avg-speed}
v_{safe} = \min \left \{ v_{max},\frac{L-C}{e1-s1} \right \}
\end{equation}
where $v_{max}$ is the speed limit on the road considered. Assuming that $\frac{L-C}{e1-s1} \leq v_{max}$, this safe speed is visualized 
in Figure \ref{fig:cp} as the slope of the blue segment connecting the points $(s1,C)$ and $(e1,L)$.
\end{appendix}

\end{document}